\documentclass[12pt,a4paper]{article}
\usepackage{mathtools}
\usepackage{mathrsfs}
\usepackage{amsmath}
\usepackage{tikz-feynman}
\usepackage[utf8x]{inputenc}
\usepackage{amssymb}
\usepackage{slashed}
\usepackage{bbm}
\usepackage{bm}
\usepackage{color}
\usepackage[a4paper,top=3cm,bottom=3cm,left=2cm,right=3cm,bindingoffset=5mm]{geometry}
\usepackage{booktabs} 
\usepackage{tikz}
\usepackage{cancel}
\usepackage[bottom]{footmisc}
\usepackage{cite}
\usepackage{float}
\usepackage{todonotes}
\usepackage{jheppub}

\usepackage{hyperref}

\hypersetup{
colorlinks=true,         
linkcolor=blue,          
citecolor=red,        
urlcolor=red            
}

\renewcommand{\vec}[1]{\mathbf{#1}}

\renewcommand{\[}{\begin{equation}\begin{aligned}}
\renewcommand{\]}{\end{aligned}\end{equation}}

\newcommand{\hel}{\eta} 
\renewcommand{\d}{\mathrm{d}}
\newcommand{\dd}{\hat{\mathrm{d}}}
\newcommand{\del}{\hat{\delta}}
\newcommand{\ket}[1]{| #1 \rangle}
\newcommand{\bra}[1]{\langle #1 |}

\renewcommand{\Re}{\operatorname{Re}}

\definecolor{allOrderBlue}{rgb}{0.4,0.5,1}

\newcommand{\gauge}{n}
\newcommand{\maxwell}{\phi}

\newcommand{\conv}{\circ}

\newcommand{\Ad}{\dot{A}}
\newcommand{\Bd}{\dot{B}}
\newcommand{\Cd}{\dot{C}}
\newcommand{\Dd}{\dot{D}}
\newcommand{\Ed}{\dot{E}}
\newcommand{\Fd}{\dot{F}}
\newcommand{\depsilon}{\epsilon}

\newcommand{\bphi}{\phi} 
\newcommand{\bB}{B} 
\newcommand{\bH}{H} 
\newcommand{\bsigma}{\sigma} 
\newcommand{\charge}{\tilde{c}} 
\newcommand{\ampA}{\mathcal{A}} 
\newcommand{\ampM}{\mathcal{M}} 

\title{NS-NS Spacetimes from Amplitudes}

\author[1]{Ricardo Monteiro,}
\author[1]{Silvia Nagy,}
\author[2]{Donal O'Connell,}
\author[1]{David Peinador Veiga,}
\author[2]{Matteo Sergola}

\affiliation[1]{Centre for Theoretical Physics, Department of Physics and Astronomy, Queen Mary University of London, E1 4NS, United Kingdom}
\affiliation[2]{Higgs Centre for Theoretical Physics, School of Physics and Astronomy, The University of Edinburgh, EH9 3FD, Scotland}

\abstract{
Recent work has shown how on-shell three-point amplitudes in gauge theory and gravity, representing the coupling to massive particles, correspond in the classical limit to the curvature spinors of linearised solutions. This connection, made explicit via the KMOC formalism in split metric signature, turns the double copy of scattering amplitudes into the double copy of classical solutions. Here, we extend this framework to the universal massless sector of supergravity, which is the complete double copy of pure gauge theory. Our extension relies on a Riemann-Cartan curvature incorporating the dilaton and the B-field. In this setting, we can determine the most general double copy arising from the product of distinct gauge theory solutions, say a dyon and $\sqrt{\text{Kerr}}$. This gives a double-copy interpretation to gravity solutions of the type Kerr-Taub-NUT-dilaton-axion. We  also discuss the extension to heterotic gravity. Finally, we describe how this formalism for the classical double copy relates to others in the literature, namely (i) why it is an on-shell momentum space analogue of the convolutional prescription, and (ii) why a straightforward prescription in position space is possible for certain vacuum solutions. 
}

\begin{document}

\begin{flushright}
QMUL-PH-21-53
\end{flushright}

\maketitle

\section{Introduction}
\label{sec:intro}

At first glance, scattering amplitudes and classical general relativity seem to have little in common. Amplitudes are quantum-mechanical objects, most 
directly relevant to phenomenology at particle colliders. Nevertheless recent years have seen an explosion of work studying applications of amplitudes
to gravity, and (conversely) the implications of deep results in gravity for scattering amplitudes, see, for example the very recent references~\cite{Herrmann:2021tct,Cristofoli:2021vyo, Bern:2020buy, Bern:2021dqo,Herrmann:2021lqe, DiVecchia:2021bdo, Bautista:2021wfy,Bern:2020gjj,Moynihan:2020gxj,Cristofoli:2020uzm,Parra-Martinez:2020dzs,Haddad:2020tvs,AccettulliHuber:2020oou,Moynihan:2020ejh,Manu:2020zxl,Sahoo:2020ryf,delaCruz:2020bbn,Bonocore:2020xuj,Mogull:2020sak,Emond:2020lwi,Cheung:2020gbf,Mougiakakos:2020laz,Carrasco:2020ywq,Kim:2020cvf,Bjerrum-Bohr:2020syg,Gonzo:2020xza,delaCruz:2020cpc,Cristofoli:2021jas,Bautista:2021llr}.

Much of this work has been motivated by the advent of gravitational-wave astrophysics \cite{Chung:2018kqs,Chung:2019duq,Bern:2020buy,Neill:2013wsa,Cheung:2018wkq,Bern:2019crd,Bern:2019nnu,Bjerrum-Bohr:2013bxa,Bjerrum-Bohr:2014zsa,Luna:2016due,Damour:2016gwp,Goldberger:2016iau,Cachazo:2017jef,Guevara:2017csg,Damour:2017zjx,Luna:2017dtq,Kosower:2018adc,Maybee:2019jus,Aoude:2021oqj,Laddha:2018rle,Laddha:2018vbn,Bjerrum-Bohr:2018xdl,Cristofoli:2019neg,Guevara:2019fsj,Kalin:2019rwq,Kalin:2019inp,Bini:2020hmy,Aoude:2020onz,Cheung:2020gyp,Cheung:2020sdj,Kalin:2020fhe,Haddad:2020que,Kalin:2020lmz,DiVecchia:2020ymx,Bern:2020uwk,AccettulliHuber:2020dal}. Like collider phenomenology, gravitational-wave physics will 
develop into a precision science --- and precision is a core success of the wider amplitudes programme. Scattering amplitudes have applications to
gravitational waves because they can be used to compute interaction potentials \cite{Chung:2018kqs,Chung:2019duq, Bern:2020buy}, or more general effective Lagrangians~\cite{Neill:2013wsa,Cheung:2018wkq,Bern:2019crd,Bern:2019nnu}, which are relevant to
compact binary coalescences. In scattering systems, amplitudes can be used rather directly to evaluate observables of interest~\cite{Kosower:2018adc,Bautista:2019tdr,Maybee:2019jus,Herrmann:2021tct,Cristofoli:2021vyo,Aoude:2021oqj}, and at least some of these
observables may then be analytically continued to the bound regime~\cite{Herrmann:2021lqe,Kalin:2019rwq,Kalin:2019inp, Bini:2020hmy}.

An additional motivation exists, however, for studying scattering amplitudes in classical gravity. This motivation is founded
on the existence of the double copy in amplitudes.

The double copy, first discovered in string theory by Kawai, Lewellen and Tye~\cite{klt} and more recently invigorated and 
deepened by Bern, Carrasco and Johansson~\cite{Bern:2008qj,Bern:2010ue}, is an algorithm for computing scattering amplitudes in gravity given amplitudes in Yang-Mills (YM) theory \cite{Bern:2008qj}.\footnote{The double copy was reviewed comprehensively in reference~\cite{Bern:2019prr}.}
From the perspective of the double copy, gravity looks to some extent like the ``square'' of YM theory. 
The double copy then provides an entirely new perspective on classical gravity, very different to the Einstein
equation and the usual geometric approach to general relativity.
Since this new perspective should make some things clear which were previously opaque, it is very interesting to study what the double copy really
means for classical spacetimes. Indeed, this has been an important topic in the recent literature, for instance, \cite{Monteiro:2014cda,Ridgway:2015fdl,White:2016jzc,Goldberger:2017frp,Adamo:2017nia,Goldberger:2017vcg,Carrillo-Gonzalez:2017iyj,Bahjat-Abbas:2017htu,Shen:2018ebu, Ilderton:2018lsf,Berman:2018hwd,Bahjat-Abbas:2018vgo,Plefka:2018dpa,Gurses:2018ckx,Luna:2018dpt,Cardoso:2016amd,Andrzejewski:2019hub,Plefka:2019hmz,CarrilloGonzalez:2019gof,Sabharwal:2019ngs,Bah:2019sda,Borsten:2019prq,Borsten:2021hua,Borsten:2021zir,Bahjat-Abbas:2020cyb,Elor:2020nqe,Adamo:2020qru,Alfonsi:2020lub,Keeler:2020rcv,Easson:2020esh,Easson:2021asd,Pasarin:2020qoa,Gumus:2020hbb,Alawadhi:2021uie,Lee:2018gxc,Berman:2020xvs,Lescano:2021ooe,Angus:2021zhy, Cheung:2021zvb,Alkac:2021seh,Ferrero:2020vww,Alkac:2021bav,Gonzo:2021drq,Saotome:2012vy,Borsten:2020zgj,Campiglia:2021srh,Godazgar:2021iae,Shi:2021qsb,Moynihan:2021rwh}.

In this article, our interest is in a systematic exploration of the double copy at its simplest: the statement that, up to a constant factor, the three point amplitude in 
gravity is the square of the three point amplitude in YM theory. This statement holds (famously) for the three-point amplitude involving three gravitons, which is the square of the three-gluon amplitude. It also holds for the three-point amplitude describing the interaction between a massive particle and 
a graviton, which is the square of the three-point amplitude describing the interaction between a charge and a gluon. In fact, we may equally well 
say that the gravitational three-point amplitude is the square of the amplitude for a photon interacting with a point charge, because at the level of these
three-point amplitudes the non-linearity of the YM field is irrelevant.

The study of amplitudes applied to classical gravity has now given us a rather direct link between amplitudes and solutions of the Einstein equation.
This link proceeds by viewing the Riemann curvature tensor far from the source as an operator in the quantum theory, whose expectation value can be computed perturbatively from
amplitudes.
This is particularly simple at linearised level, where the Riemann tensor is gauge invariant. In the correspondence regime where quantum effects are negligible, this expectation value must be the classical value of the curvature.
This idea is an application of the ``KMOC'' formalism \cite{Kosower:2018adc,Maybee:2019jus}, which was originally developed with gravitational waves in mind. From our perspective, however, 
the link allows us to directly determine what the double copy means at the level of classical solutions. 

In previous work~\cite{Monteiro:2020plf}, four of the authors studied the simplest example of this link. The photon/charge three-point amplitude computes the Coulomb field, and
maps under the double copy to the gravitational massive three-point amplitude, which computes the (linearised) Riemann curvature of the Schwarzschild
solution. Thus, we see that the double copy induces a map between Coulomb and Schwarzschild. As phrased, the map applies to linearised gravity; 
but since Schwarzschild is a Kerr-Schild spacetime, this map extends from the linearised solution to the exact solution with appropriate coordinates. 
This relation
between Coulomb and Schwarzschild is an example of the ``classical double copy'', previously proposed by two of the authors and White in~\cite{Monteiro:2014cda}. Thus, the
KMOC formalism allowed us to show (at least in this example) that the classical double copy \emph{is} the double copy.

Yet the classical double copy has been applied to solutions far more general than Coulomb and Schwarzschild. Indeed, it can be phrased more precisely at the level of curvature spinors \cite{Penrose:1960eq, Newman:1961qr,Penrose:1983mf,Penrose:1985bww}: the Weyl spinor in gravity and the Maxwell spinor in electrodynamics (i.e.~linearised YM). In that context,
the classical double copy becomes the statement that the Weyl spinor is a symmetrised square of the Maxwell spinor, up to an overall scalar factor~\cite{Luna:2018dpt}.
In this Weyl form, the classical double copy was applied to all vacuum Petrov type D solutions of the Einstein equation and has been explored also for vacuum type N 
solutions~\cite{Godazgar:2020zbv}. A twistorial interpretation of the Weyl double copy is being developed in \cite{White:2020sfn, Chacon:2021wbr,Chacon:2021hfe,Chacon:2021lox,Adamo:2021dfg,Guevara:2021yud,Farnsworth:2021wvs}.

In this article, our main goal is to connect the Weyl double copy to scattering amplitudes beyond the example in~\cite{Monteiro:2020plf}. To do so, we must consider more general amplitudes
than the simple Coulomb (charge/photon) amplitude and its double copy to a Schwarzschild (mass/graviton) amplitude. Luckily, more general amplitudes
are indeed available. In gauge theory, two non-trivial deformations of the Coulomb amplitude exist. The first of these introduces a complex phase $e^{i \eta \theta}$ in the amplitude, where $\eta$ is the helicity of the photon~\cite{Huang:2019cja}. This deformation has the interpretation of an electric/magnetic duality rotation,
allowing us to introduce magnetic charges. The second deformation is slightly more complicated. For a photon with momentum $k$, it introduces a 
factor $e^{-\eta k \cdot a}$ in the amplitude, where $a$ is a four-vector parameter. Rather remarkably, this deformation leads to an amplitude describing
a particle with large classical spin $a^\mu$ interacting with a photon. It may be derived~\cite{Arkani-Hamed:2019ymq} by studying the large spin limit of the ``minimally coupled''
amplitudes of Arkani-Hamed, Huang and Huang~\cite{ah3}, and is known to be a form of the Newman-Janis shift~\cite{Newman:1965tw}.

Indeed, in previous work~\cite{Emond:2020lwi}, these amplitudes, and their gravitational double copies, were explored in the context of the classical double copy. The double 
copies were understood as describing the Kerr-Taub-NUT family of spacetimes. However, in this work, it was not yet understood how to compute the
curvature directly from amplitudes, and so the evidence in favour of the double copy was gathered by scattering a particle off the source, and checking
that the results agree when the scattering is computed with amplitudes or the geodesic equation. Now, we are equipped to compute the curvature
directly.

Furthermore, our greater control over the relation between the amplitudes double copy and the classical double copy now allows us to explore more general
double copies, where we allow dilatons and axions to propagate in addition to gravitons. These extra degrees of freedom arise in a very simple manner.
Given amplitudes $\mathcal{A}_\pm$ for a photon of helicity $\pm$, we may construct the double copies $\mathcal{A}_+\mathcal{A}_+$ and
$\mathcal{A}_-\mathcal{A}_-$. These amplitudes describe gravitons, since the helicities add up. We can also construct, if we wish, two more amplitudes: $\mathcal{A}_+\mathcal{A}_-$ and $\mathcal{A}_-\mathcal{A}_+$. These amplitudes source a complex scalar, interpreted as a scalar dilaton and
a pseudoscalar axion. If we include these scalar amplitudes in our definition of the gravitational theory, then the full non-linear solutions
will involve an interplay between the purely gravitational and scalar degrees of freedom. 

The more general double copy at three-point level in fact connects Yang-Mills theory to the universal massless sector of supergravity, sometimes known
as NS-NS gravity. It is also known as $\mathcal{N}=0$ supergravity, although of course this is a purely bosonic theory with no supersymmetry in sight. Nevertheless, its
significant role in string theory makes this system particularly interesting. 

There is an immediate obstacle in the way of a direct application of amplitudes to solutions of NS-NS gravity. The KMOC formalism
connects curvatures to amplitudes. What kind of curvature should we use for NS-NS gravity? We find that the right object to use is a 
generalisation of the Riemann curvature associated with a torsionful, but metric-compatible, connection.  
The spinorial form of this curvature decomposes, at linearised level, into a fully symmetric (therefore spin two) object recovering the Weyl spinor, and additional objects describing the field strengths of the scalars.
As we will see, the family of solutions we encounter are of type Kerr-Taub-NUT-axion-dilaton.

It is further possible to generalise the double copy by considering asymmetric products of gauge theories: that is, to multiply two different gauge theories in the construction
of the gravitational theory. The simplest example of this idea is to multiply pure YM theory by a Yang-Mills-scalar theory. At the level of amplitudes, this
means that we may also construct amplitudes of the type $\mathcal{A}_\pm \mathcal{A}_0$, where $\mathcal{A}_0$ describes the source interacting
with a massless scalar particle. The resulting double copy theory is Einstein-Yang-Mills-axion-dilaton. These heterotic double copies are considered here for the first time in the context of the Weyl double copy, although some solutions in this theory have been studied using a double-field-theory extension of the Kerr-Schild double copy~\cite{Cho:2019ype} as well as in the convolutional approach in supersymmetric theories \cite{Cardoso:2016ngt,Cardoso:2016amd}. Interestingly, the Kerr-Newman solution has also been associated to scattering amplitudes \cite{Moynihan:2019bor}, but not as a double copy. The natural double copy theory is heterotic gravity.

Now, we turn to a crucial issue facing any application of three-point amplitudes. This issue is that, for real kinematics in Lorentzian signature, the on-shell constraints
on the three particles force the energy of the massless particle (photon, graviton or scalar) to vanish. This fact forces us to consider an analytic
continuation away from Lorentzian signature, as is familiar from the BCFW recursion relations~\cite{bcfw}. In our case, we could consider a complex contour
of integration, but as in~\cite{Monteiro:2020plf} we choose instead to work in ``split'' metric signature\footnote{For related works on split signature and its application, see also \cite{Srednyak:2013ylj,Mason:2005qu,Barrett:1993yn, Atanasov:2021oyu,Crawley:2021auj,Guevara:2021yud}.} $(+,+,-,-)$. We may analytically continue from this signature back
to the physically-relevant Lorentzian case. However, it may be worth remarking that split signature spacetimes have their own significance in general
relativity: for example, real double-Kerr-Schild coordinates are available in split signature for certain spacetimes~\cite{Plebanski:1976gy,Chong:2004}.\footnote{This requires the existence of two null vector fields that are mutually orthogonal, which is impossible in Lorentzian signature.}

Finally, we will also consider the relation of our formalism for the classical double copy, based on scattering amplitudes in momentum space, to position-space prescriptions. In particular, we will see that it provides an on-shell momentum space version of the convolutional prescription of \cite{Anastasiou:2014qba,LopesCardoso:2018xes,Anastasiou:2018rdx,Borsten:2019prq,Borsten:2020xbt,Borsten:2021zir}. Another natural question is why some solutions admit an equivalent double copy interpretation that is local in position space (i.e.~not written as a convolution), as in the Kerr-Schild double copy \cite{Monteiro:2014cda} and in the original application of the Weyl double copy \cite{Luna:2018dpt}. For various examples, we will identify the property of the solutions that allows for this simplicity. 

The structure of our paper is as follows. We will begin with a short overview which aims to give the reader a bird's eye view of the main ideas in our
paper. In section~\ref{sec:generalisedCurvature}, we explain the geometric construction of the generalised curvature relevant when axions and dilatons
are included in the double copy, thereby introducing a key tool which we will use in the remainder of the article. We move on, in section~\ref{sec:22}, to
review point charges in split signature and the determination of curvatures using amplitudes in that signature. In section~\ref{sec:YMamps}, we describe
the family of gauge solutions of interest to us. The double
copy of these solutions is the content of section~\ref{sec:doublecopy}. We work in the context of NS-NS gravity for generality. In section~\ref{sec:heterotic}, we study the asymmetric double copy case of heterotic gravity. 
Section~\ref{sec:positionSpace} deals with the Fourier transform from momentum to position space. We conclude with a discussion of our work and its implications.

\section{Overview}

The basic idea of our paper is that classical solutions can be summarised by three-point scattering amplitudes which generate them in the classical limit. These are amplitudes for the emission by a massive particle (the source) of a massless boson associated to the classical field. The amplitudes have support in Lorentzian signature with complex momenta or in split signature with real momenta. Much of the paper consists of the detailed construction of solutions from the amplitudes, i.e.~making the statements in this section explicit.

Let us start with the amplitudes in gauge theory, using Lorentzian signature for now:
\[
\text{Gauge theory:}\quad \ampA_\eta\, e^{\eta\,(-k\cdot a+i\theta)}\,, \; \eta=\pm1\,.
\label{eq:gtamp}
\]
Here, $\eta$ denotes the helicity of the gauge boson, and $k_\mu$ denotes its momentum. The case $\eta=1$ generates a self-dual field, and $\eta=-1$ an anti-self-dual field. The correspondence is
\begin{align}
& |k\rangle_A |k\rangle_B\;  \ampA_+\, e^{-k\cdot a+i\theta} \;\;\stackrel{\text{oFT}}{\longrightarrow}\;\; \phi_{AB}(x) \,,
\\
& [k|_{\dot{A}} [k|_{\dot{B}}\;  \ampA_- \,e^{k\cdot a-i\theta} \;\;  \stackrel{\text{oFT}}{\longrightarrow}\;\; \tilde{\phi}_{\dot{A}\dot{B}}(x) \,,
\end{align}
where the null momentum of the gauge boson is $k_{A\dot{B}}=|k\rangle_A [k|_{\dot{B}}$\,, and oFT stands for on-shell Fourier transform (from $k$ to $x$). The Maxwell field strength is, as usual,
\[
F_{A\dot{A}B\dot{B}}=\phi_{AB}\,\depsilon_{\dot{A}\dot{B}} + \tilde{\phi}_{\dot{A}\dot{B}}\, \epsilon_{AB}\,.
\]
Both the self-dual part and the anti-self-dual part obtained from the amplitudes are solutions to the equations of motion in their own right, but only their combination (when $\tilde{\phi}_{\dot{A}\dot{B}}$ is the complex conjugate of $\phi_{AB}$) yields a real Maxwell field in Lorentzian signature. Meanwhile, the parameters $a^\mu$ and $\theta$ in equation~\eqref{eq:gtamp} characterise the massive particle. When both vanish, the contributions from $\eta=1$ and $\eta=-1$ combine to generate the Coulomb solution, as described in \cite{Monteiro:2020plf}. The timelike vector $a^\mu$ parametrises the spin, and means that the classical `particle' is actually an extended object of size $\,a=\sqrt{|a^\mu a_\mu|}\,$ 
sourcing a dyonic gauge field. In the case $\theta=0$, this gauge field has been dubbed $\sqrt{\text{Kerr}}$, as it is the `single copy' of Kerr \cite{Monteiro:2014cda,Arkani-Hamed:2019ymq}. The dyonic parameter $\theta$ represents an electric-magnetic duality transformation \cite{Huang:2019cja}, where the cases $\,\theta=0\!\!\mod \pi\,$ and $\,\theta=\pi/2\!\!\mod \pi\,$ describe pure electric charge and pure magnetic charge, respectively.

We may generate linearised gravity solutions from the amplitudes obtained via the double copy, using `left' and `right' copies of \eqref{eq:gtamp}:
\[
\text{Gravity:}\quad \ampA_{\eta_L}\, e^{\eta_L\,(-k\cdot a_L+i\theta_L)} \times
 \ampA_{\eta_R}\, e^{\eta_R\,(-k\cdot a_R+i\theta_R)}\,.
 \label{eq:gravamp}
\]
As the gravitational analogue of the gauge field strength, we will introduce a generalised curvature tensor that incorporates all the gravity fields. In a spinorial decomposition, we have
\begin{equation}
\begin{aligned}
\mathfrak{R}_{A\Ad B\Bd C\Cd D\Dd}
&=
\mathbf{X}_{ABCD}\,\depsilon_{\Ad\Bd}\,\depsilon_{\Cd\Dd}
+\tilde{\mathbf{X}}_{\Ad\Bd\Cd\Dd}\,\epsilon_{AB}\,\epsilon_{CD}
\\
&\hspace{1cm}
+\mathbf{\Phi}_{AB\Cd\Dd}\,\depsilon_{\smash{\Ad\Bd}}\,\epsilon_{CD}
+\tilde{\mathbf{\Phi}}_{\Ad\Bd CD}\,\epsilon_{AB}\,\depsilon_{\smash{\Cd\Dd}}~.
\end{aligned}
\end{equation} 
At linearised level, the first line on the right-hand side becomes the Weyl tensor, while the second line is built from double-derivatives of a complex scalar whose real components are the dilaton and the axion. In terms of the amplitudes  \eqref{eq:gravamp}, the cases $(\eta_L,\eta_R)=(1,1)$ and $(\eta_L,\eta_R)=(-1,-1)$ correspond to self-dual and anti-self gravitons, respectively. The correspondence is
\begin{align}
& |k\rangle_A |k\rangle_B |k\rangle_C |k\rangle_D\;  (\ampA_+)^2 \, e^{-k\cdot (a_L+a_R)+i(\theta_L+\theta_R)} \;\;\stackrel{\text{oFT}}{\longrightarrow}\;\; {\mathbf{X}}_{ABCD}(x) \,,
\\
& [k|_{\dot{A}} [k|_{\dot{B}} [k|_{\dot{C}} [k|_{\dot{D}} \;  (\ampA_-)^2 \, e^{k\cdot (a_L+a_R)-i(\theta_L+\theta_R)} \;\;  \stackrel{\text{oFT}}{\longrightarrow}\;\; \tilde{\mathbf{X}}_{\dot{A}\dot{B}\dot{C}\dot{D}}(x) \,.
\end{align}
These self-dual and anti-self-dual parts combine to form the linearised Weyl tensor of the Kerr-Taub-NUT solution with rotation parameter $a_L+a_R$ and dyonic parameter $\theta_L+\theta_R$. But we also have the cases of $(\eta_L,\eta_R)=(1,-1)$ and $(\eta_L,\eta_R)=(-1,1)$, corresponding to
\begin{align}
& |k\rangle_A |k\rangle_B [k|_{\dot{C}} [k|_{\dot{D}}\;  \ampA_+\ampA_- \, e^{-k\cdot (a_L-a_R)+i(\theta_L-\theta_R)} \;\;\stackrel{\text{oFT}}{\longrightarrow}\;\; {\mathbf{\Phi}}_{AB\dot{C}\dot{D}}(x) \,,
\\
& [k|_{\dot{A}} [k|_{\dot{B}} |k\rangle_C |k\rangle_D \;  \ampA_+\ampA_- \, e^{k\cdot (a_L-a_R)-i(\theta_L-\theta_R)} \;\;  \stackrel{\text{oFT}}{\longrightarrow}\;\; \tilde{\mathbf{\Phi}}_{\dot{A}\dot{B}CD}(x) \,.
\end{align}
It is important to realise that including these extra objects is a choice: we have to decide in advance if we want our source to couple to axions and dilatons when performing the double copy.

Even if we do include the dilaton and the axion, they do not backreact on the metric at linearised level.\footnote{In the usual language of supergravity, we are defining these fields in the Einstein frame, where the propagator unambiguously separates the dilaton and axion from the gravitons.} In fact, notice that each of the four choices of $(\eta_L,\eta_R)$ generates a linearised solution to the $\mathcal N=0$ supergravity equations of motion in its own right, because the equations for the four curvature spinors decouple at linearised level. We obtain a solution from any linear superposition of the contributions from the four distinct scattering amplitudes. For instance, a solution where the only non-vanishing curvature component is ${\mathbf{X}}_{ABCD}$ is a self-dual vacuum metric. Any linearised solution can be corrected order by order in perturbation theory, leading to a fully non-linear solution. A particular solution is characterised by a particular coupling of the massive `particle' to the gravity fields. Clearly, only some linear combinations of the four $(\eta_L,\eta_R)$-solutions have a generalised curvature tensor in Lorentzian signature that is real, namely those for which ${\mathbf{X}}_{ABCD}$ and $\tilde{\mathbf{X}}_{\dot{A}\dot{B}\dot{C}\dot{D}}$ are complex conjugate, and ${\mathbf{\Phi}}_{AB\dot{C}\dot{D}}$ and $\tilde{\mathbf{\Phi}}_{\dot{A}\dot{B}CD}$ are also complex conjugate. Equivalently, we can say that the real solutions are those where $(\eta_L,\eta_R)=(\pm1,\pm1)$ contribute equally, so that the metric is real, and $(\eta_L,\eta_R)=(\pm1,\mp1)$ contribute equally, so that the dilaton and the axion are real. 

The non-uniqueness of the double copy of classical solutions is obvious in this framework. In previous work \cite{Luna:2016hge,Kim:2019jwm,Luna:2020adi}, it was argued that the most general real spacetime that can be interpreted as a double copy of the Coulomb solution is the solution discovered by Janis, Newman and Winicour (JNW) \cite{Janis:1968zz}.
The JNW solution has two parameters: mass (`graviton parameter') and dilaton parameter; Schwarzschild is the case with vanishing dilaton parameter. In the framework we present here, the two parameters arise from the linear combination of the real graviton field, such that $(\eta_L,\eta_R)=(\pm1,\pm1)$, and the real dilaton field with vanishing axion, $(\eta_L,\eta_R)=(\pm1,\mp1)$.

Notice that there is an intricate interplay of parameters in the double copy. Let us define $\bar a=a_L+a_R$ and $\Delta a=a_L-a_R$, and likewise $\bar \theta=\theta_L+\theta_R$ and $\Delta\theta =\theta_L-\theta_R$. Of these four parameters appearing in the linearised gravity solution, only the parameters $\bar a$ and $\bar \theta$ appear in the graviton components, whereas only the parameters $\Delta a$ and $\Delta \theta$ appear in the complex scalar and its conjugate.
Due to the fact that $\ampA_{+}\ampA_{-}$ is a constant (i.e.~$k$-independent), as we will review later, the complex scalar is generated by $e^{-k\cdot \Delta a+i\Delta\theta}$, and its conjugate by $e^{k\cdot \Delta a-i\Delta \theta}$. Focusing on the duality parameters $\theta$,  the effect of $\bar \theta$ is to perform a gravitational Ehlers-type `electric-magnetic' duality transformation of the metric \cite{Huang:2019cja,Alawadhi:2019urr,Banerjee:2019saj}, whereas the effect of $\Delta \theta$ is to perform an axion-dilaton duality transformation, well known from supergravity.

In fact, the double copy \eqref{eq:gravamp} is not yet fully general. That is because there is a (bi-adjoint) scalar field that is often implicit.\footnote{We will be working at linearised level, which is the starting point in the perturbative construction of the solution. Hence, just like we consider an Abelian gauge field, we consider a bi-Abelian bi-adjoint scalar.} It is useful to recall the form of the standard KLT relations for scattering amplitudes \cite{klt}: schematically, 
$$
\ampA_\text{grav}\sim \ampA_\text{YM}\times \ampA_\text{scalar}^{-1} \times\ampA_\text{YM}\,.
$$
Let us focus on three point scattering. In the original context of KLT, the three-point amplitudes involve massless states: three scalars, or three gluons, or three gravity states. Moreover, at three points, $\ampA_\text{scalar}$ is just a constant that is usually implicit. However, the relevant three-point amplitude for a linearised classical solution is that for a massive particle to emit one massless particle. So we have to characterise this massive particle in gravity, gauge theory and scalar theory (in fact, only in two of them since the double copy fixes the third one). This is the origin of further non-uniqueness in the classical double copy, which we will now illustrate. For simplicity, we will consider rotation but not a dyonic parameter, and we will restrict ourselves to real Lorentzian vacuum solutions in gravity. Hence, for gravity, we will be interested in the Kerr solution. In terms of amplitudes, the solution is generated by combining the self-dual and anti-self-dual parts,
\[
(\ampA_+)^2\, e^{-\,k\cdot a} \oplus (\ampA_-)^2 \, e^{\,k\cdot a}\;: \;\; \text{Kerr($a$)}\,,
\label{eq:Kerrdc}
\]
where $\text{Kerr($a=0$)}=\text{Schwarzschild}$. We can define the gauge theory `single copy' as
\[
\ampA_+\, e^{-k\cdot a} \oplus \ampA_-\,e^{k\cdot a}\;: \;\; \sqrt{\text{Kerr}}(a)\,,
\]
where $\sqrt{\text{Kerr}}(a=0)=\text{Coulomb}$. For the (bi-adjoint) scalar, it is natural to write
\[
e^{-k\cdot a} \oplus e^{k\cdot a}\;: \;\; \text{scalar($a$)}\,.
\]
The case\, $\text{scalar($0$)}=1/r$ \,was described in \cite{Monteiro:2020plf}: it corresponds to the scalar sourced by a static point particle. As we will see in this paper, the case with rotation \text{scalar($a$)} corresponds to the `zeroth copy' appearing as $\phi$ in the Kerr-Schild double copy for the Kerr solution \cite{Monteiro:2014cda}. The `particle' source for \text{scalar($a$)} is not exactly a point particle, but has size $a=\sqrt{|a^\mu a_\mu|}$, which is the precise counterpart to the sources for Kerr($a$) and for $\sqrt{\text{Kerr}}(a)$.\footnote{It may be surprising that, in the case of \text{scalar($a$)}, the rotating source leads to chiral amplitudes $e^{\mp k\cdot a}$ for scalar field emission. These `self-dual' and `anti-self-dual' parts of the scalar are precisely, and respectively, the complex parts appearing as $S$ and $\bar S$ in the Weyl double copy for the self-dual and anti-self-dual parts of the Weyl spinor \cite{Luna:2018dpt}. So this picture is consistent with both the Kerr-Schild double copy and the Weyl double copy.} In the choice made here, of taking the `particle' source to be naturally related in all three cases (gravity, gauge theory, scalar), the three-point KLT-like relation associated to \eqref{eq:Kerrdc} is
\[
\frac{(\ampA_+\, e^{-k\cdot a})^2}{e^{-k\cdot a}} \oplus \frac{(\ampA_-\, e^{k\cdot a})^2}{e^{k\cdot a}}\;: \;\; \text{Kerr($a$)}\,.
\label{eq:Kerrdc1}
\]

The point we want to make about the non-uniqueness arising from the bi-adjoint scalar is that \eqref{eq:Kerrdc1} is not the only possible double-copy interpretation of the Kerr solution. Another obvious possibility is
\[
\frac{(\ampA_+\, e^{-\,k\cdot a/2})^2}{1} \oplus \frac{(\ampA_-\, e^{\,k\cdot a/2})^2}{1}\;: \;\; \text{Kerr($a$)}\,.
\label{eq:Kerrdc2}
\]
Here, the single copy is
\[
\ampA_+\,e^{-k\cdot a/2} \oplus \ampA_-\, e^{k\cdot a/2}\;: \;\; \sqrt{\text{Kerr}}(a/2)\,,
\]
and the zeroth copy is
\[
1 \oplus 1\;: \;\; \text{scalar}(0)=\frac1{r}\,,
\]
which is sourced by a point particle. So the difference between the two double-copy interpretations of the Kerr solution seen here, \eqref{eq:Kerrdc1} versus \eqref{eq:Kerrdc2}, is the definition of the scalar solution, that is, of the precise nature of the massive `particle' in the three-point amplitude for the scalar. There is clearly a continuum of choices, and we have neglected above the dyonic parameter, which introduces another degree of freedom. As in the examples above, however, this freedom in the scalar just leads to different double-copy interpretations of the gravity solution, not to a new range of gravity solutions. In order to ensure the finiteness of this paper, we will not discuss all the possibilities, but hopefully the examples discussed here are clear. 

Above, we wrote down the Lorentzian three-point amplitudes. However, these are only supported on-shell for complex momenta. This means that the contour of integration for the on-shell Fourier transform (oFT) is complex. Alternatively, the amplitudes have support on real momenta in split signature. We will choose the latter option to define the oFT, and then analytically continue the solutions to Lorentzian signature. This option not only makes the oFT more transparent because the integration contour is real, but also gives us as a bonus the solutions in split signature, where, as already explored in \cite{Monteiro:2020plf}, there are causality features of interest in their own right. 
 Regarding the amplitudes above, the most important feature of the analytic continuation to split signature is the dependence on the rotation and dyonic parameters:
\[
\text{Lorentzian sign.}\;\;e^{\hel\,(-k\cdot a+i\theta)} \quad\leftrightarrow \quad \text{Split sign.} \;\;e^{\hel\,(ik\cdot a+\theta)}\,,
\]
where we take $a^\mu$ and $\theta$ to be real on both sides. We will discuss the justification in section \ref{sec:YMamps}.

In the remainder of the paper, we will present the details of the construction of solutions that is summarised in this section, illustrated by a variety of examples. We will, in addition (but not summarised here for brevity), discuss the case of heterotic gravity as a double copy, and the connection of our formalism to previous position-space prescriptions for the classical double copy.

\section{Generalised curvature and NS-NS fields}
\label{sec:generalisedCurvature}

As we have seen, the double copy of Yang-Mills theory is not only pure Einstein gravity, but more generally can be taken to be NS-NS gravity. Besides the graviton, this theory includes a scalar field $\phi$, the dilaton, and a two-form field $B_{\mu\nu}$ known as the B-field or the Kalb-Ramond field. A complete classical double copy map should include all three fields on its gravitational side. Examples of such maps have been found using double field theory, both for certain exact solutions \cite{Lee:2018gxc,Kim:2019jwm,Lescano:2021ooe,Berman:2020xvs,Angus:2021zhy} and for perturbative solutions \cite{Cho:2021nim,Diaz-Jaramillo:2021wtl}. In all these studies, the map is written in terms of fields, in contrast to the Weyl double copy, where the map relates curvatures, which are gauge invariant at the linearised level. In this section, we will address this challenge by defining a generalised curvature that packages all the NS-NS fields in geometric degrees of freedom.  We will show later that this generalised curvature tensor is the appropriate object obtained from a double copy of field strengths. 

The standard notion of geometry in general relativity, a (pseudo-)Riemanian manifold $(M,g)$ endowed with the Levi-Civita connection $\nabla$, can be generalised by relaxing the requirements on the connection. If we allow the connection to have torsion, while insisting on metric-compatibility, the result is called Riemann-Cartan geometry.

Consider a $d$-dimensional manifold $M$ equipped with a metric $g_{\mu\nu}$ and an affine connection $\mathfrak{D}$. 
In a coordinate basis, the covariant derivative acts on a vector $V$ as
\begin{equation}
\mathfrak{D}_\nu V^\mu=\partial_\nu V^\mu+\Gamma^\mu{}_{\nu\rho}\,V^{\rho}~. 
\end{equation}
In general, the affine symbols $\Gamma^\mu{}_{\nu\rho}$ do not have to be symmetric. 
Their anti-symmetric part is the \textit{torsion} tensor\footnote{See \eqref{eq:antisymmDef} for our (anti-)symmetrisation conventions, chosen for later convenience.}, $T^\mu{}_{\nu\rho}\equiv \frac{1}{2}( \Gamma^\mu{}_{\nu\rho}-\Gamma^\mu{}_{\rho\nu})=\frac{1}{2}\,\Gamma^\mu{}_{[\nu\rho]}$\,. We will take
$(M,g,\mathfrak{D})$ to be a Riemann-Cartan manifold by requiring that the connection is metric-compatible,
$$
\mathfrak{D}_\lambda\, g_{\mu\nu}=0\,.
$$
This condition constrains the affine symbols to take the form
\begin{equation}
\Gamma^\mu{}_{\nu\rho}=
\Big\{ {}^{\,\mu}_{\nu\rho}\Big\}+K^{\mu}{}_{\nu\rho}~,\label{Connection symbols}
\end{equation}
where the first term denotes the standard Christoffel symbols of the Levi-Civita connection and the second, a tensor called \emph{contorsion}, must satisfy $K_{\mu\nu\rho}=-K_{\rho\nu\mu}$\,.  
It can be written uniquely in terms of the torsion as
\begin{equation}
K^\mu{}_{\nu\rho}=
\frac{1}{2}\,g^{\mu\lambda}\left(g_{\nu\tau}\,T^{\tau}_{\ \lambda\rho}+g_{\rho\tau}\,T^{\tau}_{\ \lambda\nu}+g_{\lambda\tau}\,T^{\tau}_{\ \nu\rho}\right)~.
\end{equation}
This generalised connection defines a generalised Riemann tensor, which in our conventions we write as
\begin{equation}
\mathfrak{R}_{\mu\nu\rho}{}^\lambda=
\mathfrak{D}_\nu\Gamma^\lambda{}_{\mu\rho}
-\mathfrak{D}_\mu\Gamma^\lambda{}_{\nu\rho}
+\Gamma^\lambda{}_{\nu\tau}\Gamma^\tau{}_{\mu\rho}
-\Gamma^\lambda{}_{\mu\tau}\Gamma^\tau{}_{\nu\rho}~.
\end{equation}
It is important to note that this tensor does not have the symmetries of the usual Riemann tensor. 
It satisfies  $\mathfrak{R}_{\mu\nu\rho\sigma}=\frac{1}{2}\,\mathfrak{R}_{[\mu\nu]\rho\sigma}=\frac{1}{2}\,\mathfrak{R}_{\mu\nu[\rho\sigma]}$, but $\mathfrak{R}_{\mu\nu\rho\sigma}\neq \mathfrak{R}_{\rho\sigma\mu\nu}$ due to the lack of symmetry in the last two indices of the contorsion.
Using \eqref{Connection symbols}, it can be shown that
\begin{equation}\label{eq:FrakRasRandK}
\mathfrak{R}_{\mu\nu\rho}{}^\lambda=
R_{\mu\nu\rho}{}^\lambda
+\nabla_\nu K^\lambda{}_{\mu\rho}
-\nabla_\mu K^\lambda{}_{\nu\rho}
+K^\lambda{}_{\nu\tau}K^\tau{}_{\mu\rho}
-K^\lambda{}_{\mu\tau}K^\tau{}_{\nu\rho}~,
\end{equation}
where $\nabla$ denotes the Levi-Civita connection and $R_{\mu\nu\sigma}{}^\lambda$ its Riemann tensor. In general, $\frak{R}$ will denote curvatures with torsion, whereas $R$ is reserved for the standard Riemannian curvatures of the metric. 

Riemann-Cartan manifolds have extra geometrical degrees of freedom in the contorsion. These degrees of freedom can be used to accommodate the NS-NS fields, giving them a geometric status similar to the metric. The dilaton is assigned to the trace of the contorsion while the B-field is related to its fully antisymmetric component
\begin{gather}
K^\mu{}_{\nu\rho}=
\frac{\kappa}{2\sqrt{3}}e^{-\frac{4\kappa\phi}{d-2}}\,H^\mu{}_{\nu\rho}
-\frac{2\,\kappa}{(d-2)\sqrt{d-1}}
\left(\,
	{\delta^{\mu}}_\nu\,\partial_\rho\phi
	\,-\,g_{\nu\rho}\,g^{\mu\sigma}\,\partial_\sigma\phi
\right)~,\label{eq:ContorsionasNSNS}
\end{gather}
where $H=\d B$ is the curvature of the B-field and $\kappa$ is the gravitational coupling constant. 
The contorsion \eqref{eq:ContorsionasNSNS} was chosen such that the Ricci scalar is
\begin{equation}
\mathfrak{R}=R-\frac{4\kappa^2}{d-2}\nabla_\mu\phi\,\nabla^\mu\phi-\frac{\kappa^2}{12}e^{-\frac{8\kappa\phi}{d-2}}\,H_{\mu\nu\rho}H^{\mu\nu\rho}+\frac{4\kappa\,\sqrt{d-1}}{d-2}\nabla^\mu\nabla_\mu\phi~,\label{eq:RinTermsofFields}
\end{equation}
the motivation being that $\sqrt{|g|}\,\mathfrak{R}$ is equivalent to the usual NS-NS Lagrangian density in the Einstein frame, up to a boundary term: 
\begin{align}
S&=\frac{1}{2\kappa^2} \int\d^d x\sqrt{|g|}\left(
	R
	-\frac{4\kappa^2}{d-2}\nabla_\mu\phi\,\nabla^\mu\phi
	-\frac{\kappa^2}{12}e^{-\frac{8\kappa\phi}{d-2}}\,H_{\mu\nu\rho}H^{\mu\nu\rho}
\right)~,\\
&=\frac{1}{2\kappa^2} \int\d^d x\sqrt{|g|}\ \mathfrak{R}~.\label{eq:EHaction}
\end{align}
Similar constructions have been proposed since the discovery of the NS-NS action \cite{Scherk:1974mc}. Although the originally proposed connection was not metric compatible,  it assigned connection degrees of freedom to the dilaton and B-fields.
 This motivated a series of works trying to recast higher-order terms of the bosonic string Lagrangian exclusively in terms of generalised curvature invariants \cite{Nepomechie:1985us,Gross:1986mw,Bento:1986hx,Bern:1987wz}. A similar connection is also used in the context of double field theory \cite{Hull:2009mi}. A metric-compatible connection was introduced in \cite{Saa:1993mi}, wich together with a non-parallel volume element reproduces the NS-NS Lagrangian in the string frame. 
Other generalised connections, also metric-compatible, have been used to endow Einstein-dilaton gravity with a geometric interpretation \cite{Dunn:1974fac,Fonseca-Neto:2012cvy}.
A drawback of these geometric formulations of NS-NS gravity is that, in order to obtain the correct equations of motion, one needs to impose constraints on the torsion \cite{Dereli:1995zj}. For example, the totally antisymmetric component, which we set proportional to $H_{\mu\nu\rho}$, is not completely free, since $H$ is exact. 
Hence, the geometric interpretation of the massless modes is not entirely clear \cite{Hehl:2007bn}.

We will be interested in the curvature at linear order in the fields. Starting from $g_{\mu\nu}=\eta_{\mu\nu}+\kappa\,h_{\mu\nu}$\,, and expanding to linearised order, we obtain
\begin{equation}
\mathfrak{R}_{\mu\nu}{}^{\rho\sigma}=
-\frac{\kappa}{2}\partial_{[\mu}\partial^{[\rho}h_{\nu]}{}^{\sigma]}
+\,\frac{\kappa}{(d-2)\sqrt{d-1}}\, {\delta_{[\mu}}^{[\rho}\partial_{\nu]}\partial^{\sigma]}\phi
 +\,\frac{\kappa}{2\sqrt{3}}\partial_{[\mu}\partial^{[\rho}B_{\nu]}{}^{\sigma]}
~.
\end{equation}
In $d=4$, the field redefinitions
\[
\phi\to\frac{\sqrt{3}}{2}\bphi~,\qquad\qquad
B\to \sqrt{3}\bB~,
\]
simplify the factors to reduce the linearised Riemann tensor to
\begin{equation}
\mathfrak{R}_{\mu\nu}{}^{\rho\sigma}=
-\frac{\kappa}{2}\left(
\partial_{[\mu}\partial^{[\rho}h_{\nu]}{}^{\sigma]}
-\, {\delta_{[\mu}}^{[\rho}\partial_{\nu]}\partial^{\sigma]}\bphi
 -\partial_{[\mu}\partial^{[\rho}\bB_{\nu]}{}^{\sigma]}
 \right)\label{eq: linearised riemann}
~.
\end{equation}
This expression highlights the fact that the generalised Riemann packages all the NS-NS fields. At this order, the packaging can be taken one step further by using the `fat graviton' defined in \cite{Luna:2016hge} \footnote{Some factors differ from \cite{Luna:2016hge} due to different normalisation conventions.}
\[
\mathfrak{H}_{\mu\nu}=\mathfrak{h}_{\mu\nu}-B_{\mu\nu}-P^q_{\mu\nu}\,(2\phi+\mathfrak{h})~,
\]
where $\mathfrak{h}_{\mu\nu}$ is the trace-reversed graviton
 and $P^q_{\mu\nu}$ is a projector
\[
\mathfrak{h}_{\mu\nu}=h_{\mu\nu}-\frac{1}{2}\,h\,\eta_{\mu\nu}~,
\qquad
P^{q}_{\mu\nu}
=
\frac{1}{2}\left(\eta_{\mu\nu}-\frac{q_\mu\partial_\nu+q_\nu\partial_\mu}{q\cdot\partial}
\right)~.
\]
The constant auxiliary null vector $q^\mu$ is related to gauge choices. In fact, the terms involving $q^\mu$ drop out of the gauge-invariant curvature, which can be written as the compact expression
\[
\mathfrak{R}_{\mu\nu}{}^{\rho\sigma}=
-\frac{\kappa}{2}\,
\partial_{[\mu}\partial^{[\rho}\mathfrak{H}_{\nu]}{}^{\sigma]}~.
\]
In this sense, our generalised curvature is the `fat Riemann' associated to the `fat graviton'.

There is yet another way to rewrite \eqref{eq: linearised riemann}. In four dimensions, the two-form $\bB_{\mu\nu}$ can be traded for a pseudoscalar axion $\bsigma$, defined by
\[
\bH_{\mu\nu\rho}=-e^{2\sqrt{3}\,\bphi}\,\epsilon_{\mu\nu\rho\sigma}\partial^{\sigma}\bsigma~.
\label{eq:axiondef}
\]
At linearised order, the exponential in the expression above equals 1, and the fat Riemann is
\begin{equation}
\mathfrak{R}_{\mu\nu}{}^{\rho\sigma}=
-\frac{\kappa}{2}\left(
\partial_{[\mu}\partial^{[\rho}h_{\nu]}{}^{\sigma]}
-\, {\delta_{[\mu}}^{[\rho}\partial_{\nu]}\partial^{\sigma]}\bphi
+\epsilon^{\rho\sigma\lambda}{}_{[\mu}\partial_{\nu]}\partial_\lambda\bsigma
 \right)
~.
\end{equation}

Later in the paper, we will see how the different products of gauge theory amplitudes are associated to the different components of the generalised curvature. We will work in $d=4$, where it is convenient to use the spinor-helicity formalism for the amplitudes. The relation between the amplitudes and the generalised curvature is, therefore, much clearer if we also express the latter spinorially. As described in appendix~\ref{sec:spinorsRC}, the generalised Riemann tensor can be decomposed into spinors as 
\begin{equation}
\begin{aligned}
\mathfrak{R}_{A\Ad B\Bd C\Cd D\Dd}
&=
\mathbf{X}_{ABCD}\,\depsilon_{\Ad\Bd}\,\depsilon_{\Cd\Dd}
+\tilde{\mathbf{X}}_{\Ad\Bd\Cd\Dd}\,\epsilon_{AB}\,\epsilon_{CD}
\\
&\hspace{1cm}
+\mathbf{\Phi}_{AB\Cd\Dd}\,\depsilon_{\smash{\Ad\Bd}}\,\epsilon_{CD}
+\tilde{\mathbf{\Phi}}_{\Ad\Bd CD}\,\epsilon_{AB}\,\depsilon_{\smash{\Cd\Dd}}~,
\end{aligned}
\label{eq:curlyRspinor}
\end{equation}
where we use the bold typeface in order to distinguish the spinors from those of $R$. The spinors $\mathbf{X}_{ABCD}$, $\mathbf{\Phi}_{AB\Cd\Dd}$ and their duals are symmetric in their first and second pairs of indices.
Recall that, generically, $\mathfrak{R}_{\mu\nu\rho\sigma}\neq\mathfrak{R}_{\rho\sigma\mu\nu}$\,. 
This asymmetry implies that $\mathbf{X}_{ABCD}\neq \mathbf{X}_{CDAB}$ and $\mathbf{\Phi}_{AB\Cd\Dd}\neq \tilde{\mathbf{\Phi}}_{\Cd\Dd AB}$\,. 
The spinor $\mathbf{X}_{ABCD}$ is not completely symmetric and can be reduced further as
\begin{equation}
\begin{aligned}
\mathbf{X}_{ABCD}=\mathbf{\Psi}_{ABCD}
-\big(
\mathbf{\Sigma}_{A(C}\,\epsilon_{D)B}
+\mathbf{\Sigma}_{B(C}\,\epsilon_{D)A}
\big)
+\mathbf{\Lambda}(\epsilon_{AC}\,\epsilon_{BD}+\epsilon_{AD}\,\epsilon_{BC})
~,
\end{aligned}
\end{equation}
where $\mathbf{\Psi}_{ABCD}$ and $\mathbf{\Sigma}_{AB}$ are completely symmetric. A similar decomposition holds for $\tilde{\mathbf{X}}_{\Ad\Bd\Cd\Dd}$\,. Restricting to linearised level, we can compare the right-hand side of \eqref{eq:curlyRspinor} to the right-hand side of \eqref{eq: linearised riemann}: the first line of the former corresponds to the graviton contribution, whereas the second line corresponds to contributions from combinations of the dilaton and the axion (which is the single degree of freedom of the B-field in $d=4$). We will make this more explicit in a later section.
\\

\section{Static Point Charge in Split Signature}
\label{sec:22}
 
In this section, we will review how to extract the classical field configurations sourced by a static particle using scattering amplitudes in split signature \cite{Monteiro:2020plf}. The choice of signature is motivated by the fact that the three-point amplitude does not vanish for real kinematics in split signature. Alternatively, the same calculations could be carried out in Lorentzian signature, provided that momenta are complexified. We will address this point  more directly at the end of the section.\\

\subsection{Review}
\label{subsec:review}

Let us consider a static particle source in split signature. We will use coordinates $(t_1,t_2,x,y)$, with signature $(+,+,-,-)$.\footnote{We will also denote two dimensional space-like vectors in bold, eg. $\mathbf{x}=(x,y)$. }
 Since we have two time directions, we should specify the worldline of the particle, which we choose to be the $t_2$ axis with tangent vector $u^\mu=(0,1,0,0)$. 
This trajectory also corresponds to a static massive particle in Minkowski space, provided that one chooses to analytically continue along the $t_1$ axis; for a more in-depth discussion, see~\cite{Monteiro:2020plf}.
We will model this massive 
particle as a non-dynamical scalar wave packet, following the prescription in \cite{Kosower:2018adc}. The expectation value of the momentum of the wave packet should be $\langle p^\mu\rangle=m\,u^\mu$. We define the state as
\begin{equation}\label{state}
|\psi\rangle=\int \d\Phi(p) \, \varphi (p) \, |p\rangle, \qquad  \d\Phi(p)=\dd^4 p\, \del(p^2-m^2)\Theta(E_2) \,,
\end{equation}
where the wave function $\varphi(p)$ is sharply-peaked around the classical momentum $m\,u^\mu$. The notation $\dd$ and $\del$ packages factors of $2\pi$, as  explained in appendix \ref{sec:Conventions}. Notice that the theta function inside $\d\Phi(p)$ enforces positive energy along $t_2$, the worldline direction of the particle. The existence of the other time direction $t_1$ implies that there is another energy, $E_1$.\footnote{Note that, in the KMOC formalism, the momentum carried by messengers is of order $\hbar$, so in the classical limit our massive particle is indeed static.}

Now, we want to obtain the electromagnetic field sourced by the particle. 
The existence of two time dimensions implies that we need to specify boundary conditions for the fields. 
We choose to impose that the ``messenger'' fields (photons and gravitons) must be in a vacuum state for $t_1\to-\infty$. 
This endows the $(t_1,x,y)$ codimension-1 space with a sense of causality in which fields are sourced at $t_1=0$ by the instantaneous appearance of the particle. 
Consequently, we use the mode expansion for the gauge field operator
\[
A^\mu(x)=\sum_{\hel=\pm} \int \d\Phi(k)\,\hbar^{-\frac{1}{2}} \left( a_\hel(k)\varepsilon_\hel ^\mu(k) e^{-i\frac{k\cdot x}{\hbar}}+a^\dagger_\hel (k) \varepsilon_{\hel}^{\mu}(k) e^{i\frac{k\cdot x}{\hbar}}
\right).
\]
This time the measure is
\begin{equation}
\d\Phi(k)=\dd^4 k \, \del(k^2)\Theta(E^1) \,.
\end{equation}
Note that the theta function implies that $E_1$ must be positive. We will assume from now on that $\d\Phi(k)$ carries a $\Theta(E_1)$ for the gauge field while $\d\Phi(p)$ carries $\Theta(E_2)$ for the massive particle. The associated field strength tensor is
\begin{equation}
F^{\mu\nu}(x)=-i\sum_{\hel=\pm} \int \d\Phi(k)\hbar^{-\frac{3}{2}} \left( a_\hel(k) k^{[\mu}\varepsilon_\hel ^{\nu]} e^{-i\frac{k\cdot x}{\hbar}}-a^\dagger _\hel (k)k^{[\mu}\varepsilon_\hel ^{\nu]} e^{i\frac{k\cdot x}{\hbar}}
\right)\,.
\end{equation}
The powers of $\hbar$ will be absent in the classical limit, so from now we will set $\hbar=1$.
We want to obtain the field sourced by our particle when it is coupled to the electromagnetic field with a charge $Q$. For $t_1<0$, we impose that there must be no messengers, so the field  vanishes,
\[
\langle \psi|F^{\mu\nu}|\psi\rangle=0 \,.
\]
For positive $t_1$, the state evolves with 
\begin{equation}
|\psi_\text{out}\rangle=\lim_{t^1\to \infty}U(-t^1, t^1)|\psi\rangle=S|\psi\rangle,
\end{equation}
and the goal is to compute the expectation value of the field 
\[
\langle F^{\mu\nu} \rangle\equiv\langle \psi|S^\dagger F^{\mu\nu} S |\psi\rangle \,.
\label{eq:fexpect}
\]
Similarly, we can obtain the spinorial counterpart of the field strength tensor
\[
\maxwell_{AB}(x) = \sigma^{\mu\nu}{}_{AB} F_{\mu\nu}(x)\,.
\]
The $\sigma^{\mu\nu}$ matrices are symmetric on their spinor indices $A$ and $B$. 
These matrices project two-forms onto their self-dual parts,\footnote{In our nomenclature, a two-form $F$ is self-dual if $ F^* = F$, and anti-self-dual if $ F^* = - F$. The Hodge dual has been defined as
$
 F^*_{\mu\nu}=\frac{1}{2}\epsilon_{\mu\nu\rho\sigma}F^{\rho\sigma}~.
$} and are proportional
to the generators of SL$(2, \mathbbm{R})$.

 It was shown in \cite{Monteiro:2020plf} that, under the classical limit, the final state of the electromagnetic field is coherent:
\[
\hspace{-4pt}
S \ket{\psi} = \frac{1}{\mathcal{N}} \int \d\Phi(p) \varphi(p) \exp \left[ \sum_\hel \int \d \Phi(k)\,  \del(2 p \cdot k)\, i \ampA_{-\hel}(k) \,a^\dagger_\hel(k) \right] \ket{p} \,,
\label{eq:advert}
\]
where $\mathcal{N}$ is a normalisation factor ensuring that $\bra{\psi} S^\dagger S \ket{\psi} = 1$. For our static electric charge, the amplitude is the three-point scalar QED vertex,
\begin{equation}\label{3sqed}
\begin{split}
&\ampA_\hel(k)=-2 Q\,p\cdot \varepsilon_\hel({k}).
\end{split}
\end{equation}
The expression~\eqref{eq:advert} drastically simplifies the evaluation of expectation values. This is because the annihilation operator acts as a derivative on the state,
\[
a_\hel(k) S \ket{\psi} &= \del(2 p \cdot k) \, i\ampA_{-\hel}(k) \, S\ket{\psi} \\
&= \frac{\delta}{\delta a^\dagger_\hel(k)} S\ket{\psi} \,,\label{eq:CoherentOp}
\]
and annihilation operators can be replaced by amplitudes.
The field strength is therefore
\[
\bra{\psi} S^\dagger \, F^{\mu\nu}(x) \, S\ket{\psi} &= -2\Re i \sum_\hel \int \d\Phi(k) \, \bra{\psi} S^\dagger\, a_\hel(k) \, S \ket{\psi} \,  k^{[\mu}\varepsilon_\hel ^{\nu]} \, e^{-i k \cdot x} \\
&= \frac{1}m\Re \sum_\hel \int \d\Phi(k) \, \del(u \cdot k) \ampA_{-\hel}(k) \, k^{[\mu}\varepsilon_\hel ^{\nu]} \, e^{-i k \cdot x} \,.
\label{eq:fieldStrengthAllOrder}
\]
As shown in \cite{Monteiro:2020plf}, this can be expressed as a differential operator acting on a scalar potential $S(x)$,
\[
\langle F^{\mu\nu}(x)\rangle = Q\,u^{[\mu}\partial^{\nu]}S(x)~,
\]
with 
\begin{align}
S(x)&=2\Re\,i\int\d\Phi(k)\del(k\cdot u)e^{-ik\cdot x}~.
\label{eq:defScalarIntegral22}
\\
&=i\int \dd^4 k\, \del(k_1^2-\mathbf{k}^2)\Theta(k_1)\del(k\cdot u)\left(e^{-ik\cdot x}-e^{ik\cdot x}\right)~.\label{eq:ScalarNoticeTheta}
\end{align}

Similarly, the Maxwell spinor is
\[
\bra{\psi} S^\dagger \, \maxwell_{AB}(x) \, S \ket{\psi} &=-
\frac{\sqrt2}{m} \Re \int \d\Phi(k) \, \del(u \cdot k) \, \ket{k}_A\ket{k}_B \, e^{-i k \cdot x} \ampA_+(k),
\label{eq:ampMaxwellSpinor}
\]
while the conjugate one reads
\[
\bra{\psi} S^\dagger  \tilde \maxwell_{\Ad\Bd}(x) \, S \ket{\psi} =
\frac{\sqrt{2}}{m}\,\text{Re} \int \d\Phi ({k})\, \del(u\cdot {k}) \, [k|_{\Ad} [k|_{\Bd} \,e^{-i{k}\cdot x} \, \ampA_-(k) \,.
\label{eq:ampMaxwellConjSpinor}
\]
Thus, the two helicity amplitudes correspond directly to the two different chiralities of Maxwell spinor.

\subsection{Split signature vs Lorentzian signature}
All the classical fields we have obtained are written as integrals of three-point amplitudes over on-shell momentum space. Therefore, these integrals have no support in Lorentzian signature for real kinematics.
We have avoided this problem by using split signature. 
Alternatively, we could have proceeded in Lorentzian signature provided that we integrate over complex momenta. To illustrate these two alternatives, consider the scalar potential introduced in \eqref{eq:defScalarIntegral22}.
It can be shown that the scalar potential is related to the retarded and advanced Green's functions,
\[
S(x)=G_{\rm ret}(x)-G_{\rm adv}(x)~,
\]
where
\[
G_{\rm ret}(x)&=-\int\dd^4k\frac{e^{-i\,k\cdot x}\del(k\cdot u)}{(k_1+i\epsilon)^2-\mathbf{k}^2}
=\frac{\Theta(t_1)\Theta(t_1^2-r^2)}{2\pi\sqrt{t_1^2-r^2}}~,\\
G_{\rm adv}(x)&=-\int\dd^4k\frac{e^{-i\,k\cdot x}\del(k\cdot u)}{(k_1-i\epsilon)^2-\mathbf{k}^2}
=\frac{\Theta(-t_1)\Theta(t_1^2-r^2)}{2\pi\sqrt{t_1^2-r^2}}~.
\]
The existence of different Green's functions is linked to the freedom to choose boundary conditions. Our choice is that the field should vanish for $t_1<0$, which selects the retarded propagator. Hence, under these boundary conditions, we can write 
\[
S(x)=0 \quad \text{for} \; t_1<0\,, \quad S(x)=G_{\rm ret}(x) \quad \text{for} \; t_1>0\,.
\]

\begin{figure}
\centering
\includegraphics[scale=0.8]{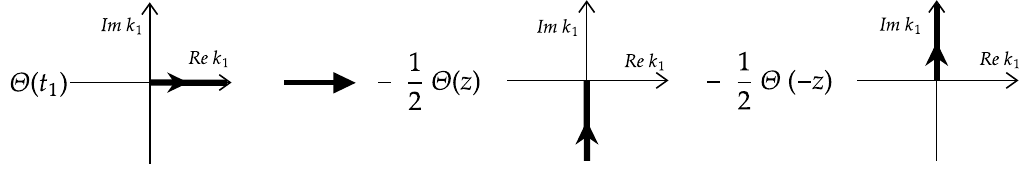}
\caption{Analytical continuation of the split signature contour to Minkowski signature.}\label{fig:contour}
\end{figure}

\begin{figure}
\centering
\includegraphics[scale=0.8]{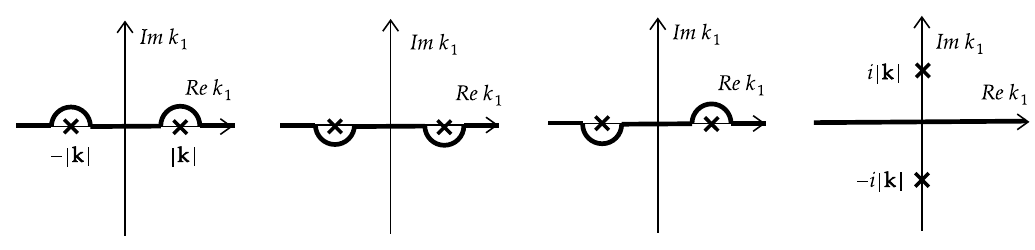}
\caption{Contour of the different Green's functions. From left to right: retarded, advanced, Feynman and the last one corresponds to Lorentzian signature.}\label{fig:poles}
\end{figure}

In Lorentzian signature, the time coordinate $t_1$ is replaced by another space coordinate, $z$, dual to $k_1$. 
Now, all three coordinates orthogonal to  $t_2$ have the same signature and no $i\epsilon$ prescription is needed. Consequently, the only Green's function is
\[
G(x)=-\int \dd^4k\,\del(k\cdot u)\,\frac{e^{-ik\cdot x}}{k^2}=\frac{1}{4\pi\sqrt{r^2+z^2}}~.
\]
Analogously to the split signature case, we can recast this Green's function into an integral of the form \eqref{eq:defScalarIntegral22}. However,  the Lorentzian delta function $\del(k_1^2+\mathbf{k}^2)$ has no roots on the real line of $k_1$. As a result, the integration contour on $k_1$ must be deformed in the complex plane. The appropriate contour is
\[
S(x)&=
-\Theta(z)\,\Re \, i \int_{-i\infty}^{0}\!\dd k_1\int \dd^2\mathbf{k}\, \del(k_1^2+\mathbf{k}^2)e^{-i k\cdot x}\\
&\hspace{2cm} -\Theta(-z)\,\Re \, i \int_{0}^{i\infty}\!\dd k_1\int \dd^2\mathbf{k}\, \del(k_1^2+\mathbf{k}^2)e^{-i k\cdot x}~.\label{eq:defScalarIntegral31}
\]
The prescription to analytically continue the retarded term in \eqref{eq:defScalarIntegral22} to Lorentzian signature is summarised in figure \ref{fig:contour}. 
The splitting or doubling of the contour might seem surprising. Ultimately, it is a reminder that the correct split signature propagator to analytically continue to Lorentzian signature is the Feynman propagator,
\[
G_{\rm F}=-\int\dd^4k\frac{e^{-i\,k\cdot x}\,\del(k\cdot u)}{k_1^2-\mathbf{k}^2+i\epsilon}
~,
\]
 which is time-symmetric. This is shown graphically in figure \ref{fig:poles}, where only the Feynman propagator contour can be deformed into the Lorentzian contour without crossing the poles.
In position space, the correct analytic continuation of the scalar potential is \footnote{This statement is explained more extensively in section 5 of \cite{Monteiro:2020plf}.}
\begin{equation}
\frac{\Theta(t_1)\Theta(t_1^2-r^2)}{2\pi\sqrt{t_1^2-r^2}}
\quad
\to
\quad
\frac{1}{4\pi\sqrt{r^2+z^2}}~.\label{eq:ContinuationScalar}
\end{equation}

Another difference in Lorentzian and split signature appears in the electromagnetic duality. In (1,3) signature, we define the self-dual and anti-self-dual electromagnetic tensors as
\[
F_{\mu\nu}^+&=\frac{1}{2}\left(F_{\mu\nu}-iF_{\mu\nu}^*\right)\,\,,\\
F_{\mu\nu}^-&=\frac{1}{2}\left(F_{\mu\nu}+iF_{\mu\nu}^*\right)~,
\]
such that $F^{\pm*}_{\mu\nu}=\pm i F^{\pm}_{\mu\nu}$. The electromagnetic stress-energy tensor can be expressed as
\[
T_{\mu}{}^\nu=F^+_{\mu\rho}\,F^{-\rho\nu}+F^-_{\mu\rho}\,F^{+\rho\nu}~.\label{eq:EnergyMomentum}
\]
Under electromagnetic duality with parameter $\theta$,
\[
F_{\mu\nu}&\to\cos\theta \,F_{\mu\nu}+\sin\theta  \,F^*_{\mu\nu}~,\\
F^*_{\mu\nu}&\to\cos\theta \,F^*_{\mu\nu}-\sin\theta  \,F_{\mu\nu}\,\,.\label{eq:DualityRot31}
\]
The self-dual and anti-self-dual tensors pick up a phase, $F^{\pm}_{\mu\nu}\to e^{\pm i\theta} F^{\pm}_{\mu\nu}$, implying that the stress-energy tensor is preserved.\\

In split signature however, the self-dual and anti-self-dual field strength tensors are
\[
F_{\mu\nu}^+&=\frac{1}{2}\left(F_{\mu\nu}+F_{\mu\nu}^*\right)~,\\
F_{\mu\nu}^-&=\frac{1}{2}\left(F_{\mu\nu}-F_{\mu\nu}^*\right)~,
\label{eq:Fplusmindef}
\]
such that $F^{\pm}_{\mu\nu}{}^*=\pm F^{\pm}_{\mu\nu}$. The stress-energy tensor is still \eqref{eq:EnergyMomentum}. In this occasion, to keep it invariant we need to have 
\[
F_{\mu\nu}&\to\cosh\theta \,F_{\mu\nu}+\sinh\theta  \,F^*_{\mu\nu}~,\\
F^*_{\mu\nu}&\to\cosh\theta \,F^*_{\mu\nu}+\sinh\theta  \,F_{\mu\nu}~,\label{eq:DualityRot22}
\]
such that $F^{\pm}_{\mu\nu}\to e^{\pm \theta} F^{\pm}_{\mu\nu}$. This difference in duality transformations \eqref{eq:DualityRot31} and \eqref{eq:DualityRot22} can be interpreted as $\theta\to-i\theta$ under analytic continuation.

\section{Gauge Fields from Amplitudes}
\label{sec:YMamps}
In section~\ref{subsec:review}, we only considered the most basic amplitude in QED for a static point particle. Now, we will generalise this amplitude to allow for magnetic charge and classical spin. The magnetic charge will be achieved by an electromagnetic duality rotation, which transforms the amplitudes as \cite{Huang:2019cja,Emond:2020lwi}
\begin{equation}\label{eq:EMrotamplitudes}
\ampA_\hel(k)
\to
\ampA_\hel(k)\,e^{\theta\,\hel}~,
\end{equation} 
Notice that the rotation parameter has been continued from Lorentzian space $\theta\to-i\theta$, as previously motivated.
 Angular momentum can be induced by a Newman-Janis shift \cite{Arkani-Hamed:2019ymq,Guevara:2020xjx}. It acts on the amplitudes as 
\begin{equation}\label{eq:NJshiftamplitudes}
\ampA_\hel(k)
\to
\ampA_\hel(k)\,e^{i\,\hel\,k\cdot a}~.
\end{equation}
The vector $a^{\mu}$ is related to the classical angular momentum. It will be taken to lie along the Wick rotated coordinate: $a^{\mu}=(a,0,0,0)$.
Consequently, the Lorentzian exponent $-\hel\,k\cdot a$ has been analytically continued to split signature as $i\,\hel\,k\cdot a$.
Neither of these transformations obstructs the exponentiation leading to the coherent state reviewed in section~\ref{subsec:review}. 
The same arguments hold, just changing the amplitudes as in \eqref{eq:EMrotamplitudes} and \eqref{eq:NJshiftamplitudes}.

The next subsections are devoted to the effects of these deformations on the field strength tensor and spinors. 
Later on, these transformations will be carried over to gravity via the double copy. 
The duality angle will be associated to the NUT charge whereas $a$ will be the Kerr angular momentum parameter.
The effect of the transformations is summarised in table \ref{tab:transformations}.
\begin{table}
\centering
\begin{minipage}{9cm}
\centering
\begin{tabular}{l|cc}
\toprule
Transformation & Gauge theory & Pure gravity \\ \midrule
None & Coulomb & Schwarzschild\\[0.7em]
EM rotation & dyon & Taub-NUT\\[0.7em]
Newman-Janis shift & $\sqrt{\text{Kerr}}$& Kerr \\[0.7em]
EM rotation $+$ NJ shift & spinning dyon & Kerr-Taub-NUT \\
\bottomrule
\end{tabular}
\caption{Effect of the transformations \eqref{eq:EMrotamplitudes} and \eqref{eq:NJshiftamplitudes}.}
\label{tab:transformations}
\end{minipage}
\end{table}
\subsection{General amplitude}
Both transformations can be applied simultaneously to the same amplitude,
\begin{equation}\label{shift}
\ampA_\eta\to \ampA_\eta e^{\,\eta\,(i k\cdot a+\theta)}~.
\end{equation}
Performing this replacement in \eqref{eq:ampMaxwellSpinor} yields the transformed Maxwell spinor 
\[
\label{eq:FTspinors}
\langle \maxwell_{AB}(x) \rangle &= -
\Re \int \d\Phi(k) \, \del(2 p \cdot k) \, 2\sqrt2\,\ket{k}_A \ket{k}_B \, \ampA_+(k) \, e^{i k\cdot a+\theta}\,e^{-i{k}\cdot x} \,\,. \\
\]
It is immediately clear that
\[
\langle \maxwell_{AB}(x) \rangle = e^\theta \langle \maxwell_{AB}^\text{Coul.}(x-a) \rangle \, , \label{ms1}
\] 
where $\langle\maxwell_{AB}^\text{Coul.}(x)\rangle$ is the Maxwell spinor of the Coulomb solution.
A similar expression can be obtained for the conjugate spinor,
\[
\langle \tilde\maxwell_{\dot A\dot B}(x) \rangle = e^{-\theta} \langle \tilde\maxwell_{\dot A\dot B}^\text{Coul.}(x+a) \rangle \, \label{ms2}
\]
The interpretation of this transformation in terms of the Newman-Janis shift and of electromagnetic duality is manifest. In the following sections, we will explore these transformations in the tensorial formalism. The advantage of the spinorial formalism will be obvious.

\subsubsection{Scalar potential}
As a preliminary step, we will study the combined effect of the transformations on the scalar potential
\[\label{scalarkern}
S_{a,\theta}(x)
&:=
2\Re i \int \d\Phi(k) \del(k\cdot u)\,e^{-i\,k\cdot(x-a)}e^{\theta}\\
&=
\frac{e^{\theta}}{2\pi}
 \,\frac{\Theta((t_1-a)^2-r^2)}{\sqrt{(t_1-a)^2-r^2}} = e^{\theta} S_{0,0}(x-a)
\]
where $r$ is defined as the $2d$ radius $\sqrt{x^2+y^2}$. The first line implies that the effect of the spin vector $a^\mu$ is merely a shift in the spacetime coordinates $x^\mu$ along $t_1$. 
The last line was obtained by introducing an $i\epsilon$ prescription for convergence, details for which can be found in appendix B of \cite{Monteiro:2020plf}. 

\subsubsection{Field strength tensor}
The classical field strength tensor can be obtained as the expectation value
\[
\langle F^{\mu\nu}(x) \rangle 
&\equiv \bra{\psi} S^\dagger \, F^{\mu\nu}(x) \, S\ket{\psi}  \\
&= -2 Q\Re \sum_\hel \int \d\Phi(k) \, \del(k \cdot u) \, e^{-i k \cdot( x+\eta\,a)} \,
k^{[\mu} \varepsilon_\hel^{\nu]} \, e^{-\theta\,\hel}\,\varepsilon_{-\hel} \cdot u \, .
\]
In the second line, we have substituted the amplitude into \eqref{eq:fieldStrengthAllOrder}. The integrand can be expanded as
\begin{multline}
\langle F^{\mu\nu}(x) \rangle 
=
 -2 Q\Re  \int \d\Phi(k) \, \del(k \cdot u) \, e^{-i k \cdot x} \,
\\\times
\left(
k^{[\mu} \varepsilon_+^{\nu]} \, e^{-\theta-i\,k\cdot a}\,\varepsilon_{-} \cdot u
+
k^{[\mu} \varepsilon_-^{\nu]} \, e^{\theta+i\,k\cdot a}\,\varepsilon_{+} \cdot u
\right)
 \, .
\end{multline}
Using the null tetrad
\[
\eta^{\mu\nu}=k^{(\mu}\gauge^{\nu)}-\varepsilon_+^{(\mu}\varepsilon_{-}^{\nu)}~,
\qquad k\cdot n=1~,
\label{eq:tetrad}
\]
the tangent vector can be decomposed into
\[
u^\mu = (u \cdot \gauge) \, k^\mu - \varepsilon_- \cdot u \, \varepsilon_+^\mu - \varepsilon_+ \cdot u \, \varepsilon_-^\mu \,\,,
\label{eq:velocityExpansion}
\]
which implies that the above expression can be rearranged as
\[
\langle F^{\mu\nu}(x) \rangle 
&=
 2 Q\Re  \int \d\Phi(k) \, \del(k \cdot u) \, e^{-i k \cdot x} \\
&\times\left(
	\cos(k\cdot a-i\theta)\,k^{[\mu}u^{\nu]}
	+{i\,\sin(k\cdot a-i\theta)}
	\left(
		k^{[\mu} \varepsilon_+^{\nu]} \,\varepsilon_{-} \cdot u
		-
		k^{[\mu} \varepsilon_-^{\nu]} \, \varepsilon_{+} \cdot u
	\right)
\right)
 \, .
\]
This can be further simplified to
\[\label{Fmunu}
\langle F^{\mu\nu}(x) \rangle 
&=
 2 Q\Re  \int \d\Phi(k) \, \del(k \cdot u) \, e^{-i k \cdot x} \\
&\times
\left(
	\cos(k\cdot a-i\theta)\,k^{[\mu}u^{\nu]}-\frac{i\,\sin(k\cdot a-i\theta)}{2}
	\epsilon^{\mu\nu\rho\sigma}k_{[\rho}u_{\sigma]}
\right),
 \, 
\]
using the result of the equation \eqref{eq:epsku}.
Expanding the sine and cosine and making use of the previously defined $S_{a,\theta}(x)$, we obtain
\[
\langle F^{\mu\nu}(x) \rangle 
&=Q\,\partial^{[\mu}u^{\nu] }
\left[
	\frac{S_{a,\theta}(x)+S_{-a,-\theta}(x)}{2}
\right]
-\frac{Q}{2}\epsilon^{\mu\nu\rho\sigma}\partial_{[\rho}u_{\sigma] }
\left[
	\frac{S_{a,\theta}(x)-S_{-a,-\theta}(x)}{2}
\right].
\]
In this expression, the derivatives act on the terms in brackets, since $u^\mu$ has constant components.
We have obtained the fields for generic $a$ and $\theta$. In the next subsections, we will focus on each transformation individually and interpret their effects. 

\subsection{Newman-Janis shift}
A pure Newman-Janis shift has $\theta=0$,
\[
S_{a,0}(x)
&=
\frac{1}{2\pi}
\,\frac{\Theta\left(
	(t_1-a)^2-r^2
\right)}{\sqrt{(t_1-a)^2-r^2}}~.
\label{eq:S_a0}
\]
It is worth remarking on some features of this field. First of all, it encodes the complex zeroth copy of Kerr. This can be checked by rotating back to (1,3) signature as $t_1\to i\,z$ 
\[
(t_1-a)^2-r^2
\to
(i\,z-a)^2-r^2
=
-(z+ia)^2-r^2
=
-\left(\tilde R+i\,a\,\cos\vartheta\right)^2~.
\]
The ``Kerr radius'' $\tilde R$ and polar angle $\vartheta$ are implicitly defined by
\begin{equation}
\frac{r^2}{\tilde R^2+a^2}+\frac{z^2}{\tilde R^2}=1~,\qquad\cos\vartheta=\frac{z}{\tilde R}~.
\end{equation}
 Thus,
\begin{equation}
\frac{1}{\sqrt{(t_1-a)^2-r^2}}
\quad\to\quad
\frac{i}{\tilde R+i\,a\,\cos\vartheta}\,
\end{equation}
which is proportional to the scalar (3.32) in \cite{Luna:2018dpt}. The fact that we recover the complex scalar supports the idea of associating the double copy for amplitudes in (2,2) signature to the Weyl double copy. 
Secondly, notice that the spin $a$ appears only in the combination $(t_1 - a)$ in equation~\eqref{eq:S_a0}. This is the Newman-Janis shift
at work: in (2,2) signature, the shift is a \emph{real} translation, in the $t_1$ direction, at the level of the field strength. (The shift acts in a 
more subtle way on the potential. At the level of the effective action, the shift can be interpreted as replacing the usual worldline action
with a worldsheet structure~\cite{Guevara:2020xjx}.)

Setting $\theta=0$ in the field strength tensor yields
\[
\langle F^{\mu\nu}(x) \rangle 
&=Q\,\partial^{[\mu}u^{\nu] }
\left[
	\frac{S_{a,0}(x)+S_{-a,0}(x)}{2}
\right]
-\frac{Q}{2}\epsilon^{\mu\nu\rho\sigma}\partial_{[\rho}u_{\sigma] }
\left[
	\frac{S_{a,0}(x)-S_{-a,0}(x)}{2}
\right].\label{eq:sqrKerrF}
\]
Note that in Minkowski signature the first bracket corresponds to the real part of $S$ while the second corresponds to the imaginary part. In
this formulation, the interpretation of the sign of the Newman-Janis translation looks somewhat complicated, compared to the Maxwell spinors seen above.

The field \eqref{eq:sqrKerrF} is the split signature equivalent of the $\sqrt{\text{Kerr}}$ solution \cite{Monteiro:2014cda,Arkani-Hamed:2019ymq}. 
Instead of checking this claim by direct comparison, which would be tedious due to coordinate transformations, we can derive \eqref{Fmunu} with $\theta=0$ in a purely classical way.
To do so, we solve the Maxwell equations in the presence of a $\sqrt{\text{Kerr}}$ source, 
\begin{equation}\label{eomrk}
\begin{split}
\partial_\mu F^{\mu\nu}(x)=Q \int \d\tau\,  {\exp( a*\partial)^\nu}_\rho u^\rho \delta^{(4)}(x-u\tau)\equiv j^\nu_{\sqrt{\text{Kerr}}}(x),\,\,\,\,\, (a*b)_{\mu\nu}=\epsilon_{\mu\nu\rho\sigma}a^\rho b^\sigma.
\end{split}
\end{equation}
Note that the source for $\sqrt{\text{Kerr}}$ is formally the one for Coulomb (see for instance \cite{justin2017} or \cite{Guevara:2020xjx}) but acted upon by the differential operator ${\exp( a*\partial)^\nu}_\rho $.
We observe that the exponential $\exp(a*\partial)$ remains invariant under analytic continuation to split signature. This is because the derivative picks up a factor of $-i$, which is cancelled out by the $i$ picked up by the volume form.

We solve  \eqref{eomrk} by Fourier transform with the boundary conditions outlined in \ref{subsec:review}. We get 
\begin{equation}\label{AAA}
\begin{split}
A^\mu(x)& = 2Q\,  \text{Re}\, i\! \int {\d} \Phi(k) \,\hat{\delta}(k\cdot u)e^{-ik\cdot x} {\exp( -ia*k)^\mu}_\rho u ^\rho.
\end{split}
\end{equation}
The action of the exponential matrix can be further simplified. In fact, on the support of the on-shell measure, it can be shown that
\begin{equation}\label{eul}
 {\exp( -ia*k)^\mu}_\rho u ^\rho=  u^\mu \cos a\cdot k-i \epsilon^\mu (a, k, u) \frac{\sin a\cdot k}{a\cdot k},
\end{equation}
where we defined $\epsilon^\mu(a, b,c ):=\epsilon^{\mu\alpha\beta\gamma}a_\alpha b_\beta c_\gamma$. We obtain finally
\begin{equation}
A^\mu(x)= 2Q\,  \text{Re}\, i\! \int {\d} \Phi(k) \,\hat{\delta}(k\cdot u)e^{-ik\cdot x} \left( u^\mu \cos a\cdot k-i \epsilon^\mu (a, k, u) \frac{\sin a\cdot k}{a\cdot k}\right).
\end{equation}
The Maxwell tensor is then easily computed,
\begin{equation}\label{Fa}
F^{\mu\nu}(x)= 2 Q\,  \text{Re}\, \! \int {\d} \Phi(k) \,\hat{\delta}(k\cdot u)e^{-ik\cdot x} \left(
k^{[\mu}u^{\nu]}\cos a\cdot k -\frac{i}{2}
\epsilon^{\mu\nu \rho\sigma}k_{[\rho}u_{\sigma]} \sin a\cdot k
\right),
\end{equation}
which is equal to the $F_{\mu\nu}$ we had in the purely spinning case with $\theta=0$.\\

Furthermore, starting from \eqref{Fa} we can also confirm  the expressions \eqref{ms1} and \eqref{ms2} in the  $\theta=0$ case. Projecting on a spinor basis, these are found to be
\begin{equation}\label{maxspinor}
\begin{split}
\phi^{\sqrt{\text{Kerr}}}_{AB} (x)&={{\sigma}^{\mu\nu}}_{AB}F_{\mu\nu}(x)\\&=  2\sqrt{2} Q\,  \text{Re}\, \! \int {\d} \Phi(k) \hat{\delta}(k\cdot u)e^{-ik\cdot (x-a)}\varepsilon_+\cdot u \,|k \rangle_A |k\rangle_B\\&=\phi^{\text{Coul.}}_{AB}(x-a),
\end{split}
\end{equation}
we  report the negative-helicity spinor too
 \begin{equation}
 \begin{split}
\tilde{\phi}^{\sqrt{\text{Kerr}}}_{\Ad\Bd} (x)&={\tilde{\sigma}^{\mu\nu}}_{\,\,\,\,\,\,\Ad\Bd}F_{\mu\nu}(x)\\&=  -2\sqrt{2} Q\,  \text{Re}\, \! \int {\d} \Phi(k) \hat{\delta}(k\cdot u)e^{-ik\cdot (x+a)}\varepsilon_-\cdot u \,[k|_{\Ad}  [k|_{\Bd}\\&=\tilde{\phi}^{{\text{Coul.}}}_{\Ad\Bd}(x+a),
\end{split}
\end{equation}
matching the expressions first obtained in \cite{Guevara:2020xjx}.

Notice again that the action of the Newman-Janis translation on the Maxwell spinors is beautifully simple: $\phi^{\sqrt{\text{Kerr}}}_{AB}$ is
a translation of $\phi^{\text{Coul.}}_{AB}$ in one direction, while $\tilde{\phi}^{\sqrt{\text{Kerr}}}_{\Ad\Bd}$ is a translation of $\tilde{\phi}^{{\text{Coul.}}}_{\Ad\Bd}$ in the opposite direction. This is in contrast to the more complicated structure at the level of the field strength~\eqref{eq:sqrKerrF}. We see that the notion of chirality is intimately related to the structure of the Newman-Janis shift.

\subsection{Duality rotation}
\label{sec:dyon}
To investigate the effect of the EM rotation on the fields, we take $a$ to zero,
\[
\langle F^{\mu\nu}(x) \rangle 
&=Q\,\partial^{[\mu}u^{\nu] }
\left[
	\frac{S_{0,\theta}(x)+S_{0,-\theta}(x)}{2}
\right]
-\frac{Q}{2}\epsilon^{\mu\nu\rho\sigma}\partial_{[\rho}u_{\sigma] }
\left[
	\frac{S_{0,\theta}(x)-S_{0,-\theta}(x)}{2}
\right]\,.
\]
Substituting the value of the scalar integrals,
\[
\langle F^{\mu\nu}(x) \rangle 
&=\cosh\theta\,Q\,\partial^{[\mu}u^{\nu] }
\left[
	\frac{\Theta(\rho^2)}{2\pi\,\rho}
\right]
-\sinh\theta\,\frac{Q}{2}\epsilon^{\mu\nu\rho\sigma}\partial_{[\rho}u_{\sigma] }
\left[
	\frac{\Theta(\rho^2)}{2\pi \rho}
\right]~,\label{eq:EMrotatedF}
\]
where $\rho^2\equiv x^2-(x\cdot u)^2$. 
In the region $\rho^2>0$, we can Wick rotate back to Lorentzian signature. The duality angle transforms as $\theta \to i \theta$ and so $\cosh\theta\to\cos \theta$ and $\sinh\theta\to -i\sin\theta$. This factor of $i$ is absorbed by the continuation of the volume form to Lorentzian signature, yielding 
\begin{equation}
\langle F^{\mu\nu} \rangle =\cos\theta\, F^{\mu\nu}_{\text{Coul.}}+\sin\theta\, F^{*\mu\nu}_{\text{Coul.}}~, \label{eq:EMrotatedF31}
\end{equation}
where, again, the star denotes Hodge conjugation. We conclude that the transformation \eqref{eq:EMrotamplitudes} generates a dyon by a duality rotation of the field.

\section{NS-NS fields from Amplitudes}
\label{sec:doublecopy}
Motivated by the double copy of amplitudes, we will consider the map
\[\label{eq:DCmap}
\ampM_{\hel_L \hel_R}
=
-\frac{\kappa}{4\,Q^2}\,c_{\hel_L \hel_R}\,\ampA^{(L)}_{\hel_L}\,\ampA^{(R)}_{\hel_R}~,
\] 
where there are four choices for $(\hel_L, \hel_R)$:
\[
(+,+)\,,\quad (-,-)\,,\quad (+,-)\,,\quad (-,+)\,.
\]
These correspond, respectively, to the gravity field being: positive-helicity graviton, negative-helicity graviton, complex scalar (dilaton and axion), and conjugate complex scalar. In general, we allow for four distinct couplings $c_{\hel_L \hel_R}$ of our massive particle to these gravity fields. Any choice of these couplings will lead to a linearised gravity solution. In practice, we will be most interested in the case where the particle couples equally to the two chiralities, in which case we take $c_{++}=c_{--}$ and $c_{+-}=c_{-+}$.

Consider the following mode expansion of the fat Riemann operator
\begin{equation}
{\mathfrak{R}}^{\mu\nu\rho\sigma}
=\kappa\,\Re
\int \d\Phi(k)
\bigg[
\sum_{\hel_L \hel_R}\,a_{\hel_L \hel_R}\, \varepsilon_{\hel_L}^{[\mu}(k)k^{\nu]}\varepsilon_{\hel_R}^{[\rho}(k)k^{\sigma]}
\bigg]e^{-ik\cdot x}~.\label{eq: linearised riemann operator}
\end{equation}
The operator version of the linearised spinor coefficients are computed by contracting with the sigma matrices \cite{Penrose:1983mf}
\begin{align}
\mathbf{X}_{ABCD}
&= 
{\sigma^{\mu\nu}}_{AB}{\sigma^{\rho\sigma}}_{CD}\,\mathfrak{R}_{\mu\nu\rho\sigma}~,
\qquad
\tilde{\mathbf{X}}_{\Ad\Bd\Cd\Dd}
= 
{\tilde\sigma^{\mu\nu}}_{\,\,\,\,\,\Ad\Bd}{\tilde\sigma^{\rho\sigma}}_{\,\,\,\,\,\Cd\Dd}\,\mathfrak{R}_{\mu\nu\rho\sigma}~,
\\[0.7em]
\mathbf{\Phi}_{AB\Cd\Dd} 
&=
{\sigma^{\mu\nu}}_{AB}{\tilde\sigma^{\rho\sigma}}_{\,\,\,\,\,\Cd\Dd}\,\mathfrak{R}_{\mu\nu\rho\sigma}~,
\qquad
\mathbf{\tilde\Phi}_{\Ad\Bd CD} 
=
{\tilde{\sigma}^{\mu\nu}}_{\,\,\,\,\, \Ad \Bd}{{\sigma}^{\rho\sigma}}_{CD}\,\mathfrak{R}_{\mu\nu\rho\sigma}~.
\end{align}
These contractions are easily computed applying 
\[
{\sigma_{\mu\nu}}^{AB} \, k^{[\mu} \varepsilon_-^{\nu]} &=- \sqrt{2} \, \bra{k}^A \bra{k}^B \,, \\
{\tilde \sigma_{\mu\nu}}^{\,\,\,\,\,\,\Ad\Bd} \, k^{[\mu} \varepsilon_-^{\nu]} &= 0 \,,
\label{eq:negativeHelFS}
\]
\[
{\sigma_{\mu\nu}}^{AB} \, k^{[\mu} \varepsilon_+^{\nu]} &= 0 \,, \\
{\tilde{\sigma}_{\mu\nu}}^{\,\,\,\,\,\,\Ad\Bd} \, k^{[\mu} \varepsilon_+^{\nu]} &=  \sqrt{2} \, |k]^{\Ad} |k]^{\Bd} \,.
\label{eq:positiveHelFS}
\]
The resulting spinors are
\begin{align}
{\mathbf{X}}_{ABCD}
&=
\kappa\,\Re\,2\int\d \Phi(k)\; a_{--} \ket{k}_A\ket{k}_B
\ket{k}_C\ket{k}_D\,e^{-ik\cdot x}~,
\\[1em]
\tilde{\mathbf{X}}_{\Ad\Bd\Cd\Dd}
&=
\kappa\,\Re\,2\int\d \Phi(k)\; a_{++} [k|_{\Ad}[k|_{\Bd}
[k|_{\Cd}[k|_{\Dd}\,e^{-ik\cdot x}~,
\\[1em]
{\mathbf{\Phi}}_{AB\Cd\Dd}
&=
-\kappa\,\Re 2\int\d\Phi(k)\; a_{-+}
\,\ket{k}_A\ket{k}_B[k|_{\Cd}[k|_{\Dd}\,e^{-ik\cdot x}~,
\\[1em]
{\tilde{ \mathbf{\Phi}}}_{\Ad\Bd CD}
&=
-\kappa\,\Re 2\int\d\Phi(k)\; a_{+-}
\,[k|_{\Ad}[k|_{\Bd} \ket{k}_C\ket{k}_D\,e^{-ik\cdot x}~.
\end{align}

In order to link these objects to the amplitudes \eqref{eq:DCmap}, we would like to use the equivalent of  \eqref{eq:CoherentOp} for the NS-NS fields. Appendix \ref{sec:CoherentState} shows that the gravitational final state is also coherent, so
\[
\hspace{-4pt}
S \ket{\psi} 
&=
 \frac{1}{\mathcal{N}} \int \d\Phi(p) \varphi(p)\\
&\hspace{40pt}
\exp \bigg[ 
	 \int \d \Phi(k)\,  
	i\,\del(2 p \cdot k)\,
	\left(
		\sum_{\hel_L \hel_R} \ampM_{-\hel_L ,-\hel_R}(k) \,a^\dagger_{\hel_L \hel_R}(k) 
	\right)
\bigg] \ket{p} \,,
\label{eq:NSNSCoherent1}
\]
which is analogous to \eqref{eq:advert}. Hence, we conclude that 
\[
a_{\hel_L \hel_R}(k) S \ket{\psi} &= \del(2 p \cdot k) \, i\ampM_{-\hel_L,-\hel_R}(k) \, S\ket{\psi} \\
&= \frac{\delta}{\delta a^\dagger_{\hel_L\hel_R}(k)} S\ket{\psi} \,.\label{eq:CoherentNSNSOp}
\]
\\
Equation \eqref{eq:CoherentNSNSOp} implies that we can easily exchange annihilation operators for amplitudes inside expectation values, so that we find
\begin{multline} 
\langle{\mathfrak{R}}^{\mu\nu\rho\sigma}\rangle
=\kappa\,\Re\, i
\int \d\Phi(k)\del(2k\cdot p)
\bigg[
\sum_\eta\,\ampM_{-\hel_L,- \hel_R}\, \varepsilon_{\hel_L}^{[\mu}(k)k^{\nu]}\varepsilon_{\hel_R}^{[\rho}(k)k^{\sigma]}
\bigg]e^{-ik\cdot x}~.\label{eq:lienarRiemannExpectation}
\end{multline}
The same can be done in the spinor coefficients. The application of the map \eqref{eq:DCmap} results in
\begin{align}
\langle{\mathbf{X}}_{ABCD}\rangle
&=
-\frac{\kappa^2c_{++}}{2Q^2}\,\Re\,i\int\d \Phi(k)  \del(2 p \cdot k) \,
	\ampA_+^{(L)}\ampA_+^{(R)}\,
\ket{k}_A\ket{k}_B
\ket{k}_C\ket{k}_D\,e^{-ik\cdot x}~,
\\[1em]
\langle\tilde{\mathbf{X}}_{\Ad\Bd\Cd\Dd}\rangle
&=
-\frac{\kappa^2c_{--}}{2Q^2}\,\Re\,i\,\int\d \Phi(k) \del(2 p \cdot k) 
 \,
	\ampA_-^{(L)}\ampA_-^{(R)}\,
[k|_{\Ad} [k|_{\Bd}
[k|_{\Cd}[k|_{\Dd}\,e^{-ik\cdot x}~,
\\[1em]
\langle{\mathbf{\Phi}}_{AB\Cd\Dd}\rangle
&=
+\frac{\kappa^2c_{+-}}{2\,Q^2}\,\Re i\int\d\Phi(k) \del(2 p \cdot k) 
 \,
	\ampA_+^{(L)}\ampA_-^{(R)}\,
\,\ket{k}_A\ket{k}_B[k|_{\Cd}[k|_{\Dd}\,e^{-ik\cdot x}~,
\\[1em]
\langle{\tilde{ \mathbf{\Phi}}}_{\Ad\Bd CD}\rangle
&=
+\frac{\kappa^2c_{-+}}{2\,Q^2}\,\Re i\int\d\Phi(k) \del(2 p \cdot k) 
 \,
	\ampA_-^{(L)}\ampA_+^{(R)}\,
\,[k|_{\Ad}[k|_{\Bd} \ket{k}_C\ket{k}_D\,e^{-ik\cdot x}~.
\end{align}
The above expressions make it clear that every quadratic term in the amplitudes sources a different component of the spinorial curvature. 
The double copy structure is remarkably explicit when comparing with the gauge field spinors \eqref{eq:ampMaxwellSpinor} and \eqref{eq:ampMaxwellConjSpinor}. 
Moreover, we can obtain the equivalent of \eqref{ms1} and \eqref{ms2}
by considering the gauge amplitudes
\[
\ampA^{(L)}_\eta=-2Q(p\cdot\varepsilon_\eta)e^{\hel(\theta_L+ik\cdot a_L)},
\\
\ampA^{(R)}_\eta=-2Q(p\cdot\varepsilon_\eta)e^{\hel(\theta_R+ik\cdot a_R)},
\]
which under the double copy map imply that 
\[
\label{eq:finaleq}
\langle{\mathbf{X}}_{ABCD}(x)\rangle
&=e^{\bar{\theta}}\langle{\mathbf{X}}_{ABCD}^\text{JNW}(x-\bar{a})\rangle~,
\\[1em]
\langle\tilde{\mathbf{X}}_{\Ad\Bd\Cd\Dd}(x)\rangle
&=e^{-\bar{\theta}}\langle\tilde{\mathbf{X}}_{\Ad\Bd\Cd\Dd}^\text{JNW}(x+\bar{a})\rangle~,
\\[1em]
\langle{\mathbf{\Phi}}_{AB\Cd\Dd}(x)\rangle
&=e^{\Delta\theta}\langle{\mathbf{\Phi}}_{AB\Cd\Dd}^\text{JNW}(x-\Delta a)\rangle~,
\\[1em]
\langle{\tilde{ \mathbf{\Phi}}}_{\Ad\Bd CD}(x)\rangle
&=e^{-\Delta \theta}\langle{\tilde{ \mathbf{\Phi}}}_{\Ad\Bd CD}^\text{JNW}(x+\Delta a)\rangle~,
\]
where we have defined
\begin{equation}\label{angledef}
\bar{\theta}:= \theta_L+\theta_R~,
\qquad\Delta\theta:=\theta_L-\theta_R~,
\end{equation}
\begin{equation}
\bar{a}:=a_L+a_R, \qquad \Delta a:=a_L-a_R \,.
\end{equation}
The superscript JNW refers to the solution where both single copies are Coulomb. Notice that, at linearised level, the first two spinors in \eqref{eq:finaleq} match those of the Schwarzschild solution. The various parameters are elegantly distributed over the different spinors. The parameter $\bar a$ corresponds to the spin of Kerr, and appears as expected via the Newman-Janis shift, while $\bar \theta$ corresponds to the split-signature version of the rotation between the mass and the NUT parameter; together, these two parameters correspond to the Kerr-Taub-NUT solution. The parameters $\Delta a$ and $\Delta \theta$ correspond, respectively, to a novel type of Newman-Janis  shift for the axion and dilaton, and to the standard axion-dilaton supergravity duality transformation. While this draft was in preparation, the Kerr and Kerr-Taub-NUT solutions were discussed, also in split signature, in~\cite{Crawley:2021auj,Guevara:2021yud}.

The spinorial language is better fitted for displaying the double copy, but it is instructive to think about the dilaton and axion. 
We can map \eqref{eq:lienarRiemannExpectation} to the field degrees of freedom using \eqref{eq: linearised riemann}, together with the mode expansions of the fields
\begin{equation}
h^{\mu\nu}=2\,\Re\sum_\eta\int \d\Phi(k)a_{\eta\eta}(k)\varepsilon^\mu_\eta(k)\varepsilon^\nu_\eta(k)\,e^{-ik\cdot x}~,
\end{equation}
\begin{equation}
\bphi=2\,\Re\int \d\Phi(k)\, a_{\bphi}(k)\,e^{-ik\cdot x}~,\label{eq:DilatonOp}
\end{equation}
\begin{equation}
\bB^{\mu\nu}=2\,\Re\int \d\Phi(k)\, a_{\bB}(k)
\left(
	\varepsilon^{\mu}_+(k)\varepsilon^{\nu}_-(k)
	-\varepsilon^{\mu}_-(k)\varepsilon^{\nu}_+(k)
\right)	
\,e^{-ik\cdot x}~.\label{eq:BfieldOp}
\end{equation} 
Substituting in \eqref{eq: linearised riemann} implies that \eqref{eq:lienarRiemannExpectation} can be re-expressed as
\begin{multline}
{\mathfrak{R}}^{\mu\nu\rho\sigma}
=\kappa\,\Re
\int \d\Phi(k)
\bigg[
\sum_\eta\,a_\eta\varepsilon_\eta^{[\mu}(k)k^{\nu]}\varepsilon_\eta^{[\rho}(k)k^{\sigma]}
\\
+a_{\bphi}\, k^{[\mu}\eta^{\nu][\rho}k^{\sigma]}
+a_{\bB}\, k^{[\mu}(\varepsilon_+^{\nu]}\varepsilon_-^{[\rho}-\varepsilon_-^{\nu]}\varepsilon_+^{[\rho})k^{\sigma]}
\bigg]e^{-ik\cdot x}~.\label{eq: linearised riemann operator fields}
\end{multline}
The first term in the second line of \eqref{eq: linearised riemann operator fields} needs simplification. This is achieved by expanding the flat metric in terms of the null tetrad
\[
k^{[\mu} \eta^{\nu][\rho} k^{\sigma]} 
=
 -k^{[\mu} \varepsilon_+^{\nu]} \varepsilon_-^{[\rho} k^{\sigma]} -k^{[\mu} \varepsilon_-^{\nu]} \varepsilon_+^{[\rho} k^{\sigma]} \,.\label{eq:antisymketak}
\]
Comparison to \eqref{eq: linearised riemann operator} then implies the following relations between annihilation operators
\[
a_{++}=a_+~,
\qquad \qquad a_{-+}=a_{\bphi}+a_{\bB} ~,\\
a_{--}=a_-~,
\qquad \qquad
a_{-+}=a_{\bphi}-a_{\bB}~,
\]
and hence the corresponding relation for amplitudes
\[\label{eq:MixedAmplitudes}
\ampM_{\hel\hel}
&=
-\frac{\kappa}{4\,Q^2}\,c_{\hel\hel}\,\ampA^{(L)}_\hel\,\ampA^{(R)}_\hel
~,\\
\ampM_\phi
&=
-\frac{\kappa}{4\,Q^2}\;\frac{1}{2}\left(c_{+-} \ampA^{(L)}_+\ampA^{(R)}_-+ c_{-+}\ampA^{(L)}_-\ampA^{(R)}_+\right)
~,\\
\ampM_B
&=
-\frac{\kappa}{4\,Q^2}\;\frac{1}{2}\,\left(c_{+-} \ampA^{(L)}_+\ampA^{(R)}_-- c_{-+} \ampA^{(L)}_- \ampA^{(R)}_+\right)~.
\] 
In the next sections this prescription will be put into practice to compute the classical fields obtained by double copying the amplitudes discussed in section \ref{sec:YMamps}. In the following we will restrict to the case $c_{++}=c_{--}$ and $c_{+-}=c_{-+}$ since these solutions
naturally continue to real solutions in Minkowski signature.

\subsection{Duality rotation}\label{sec:TaubNUT}
We will now turn to a concrete example. Consider left and right amplitudes that differ in their EM duality angle,
\[
\ampA^{(L)}_\eta=-2Q(p\cdot\varepsilon_\eta)e^{\theta_L\eta},
\\
\ampA^{(R)}_\eta=-2Q(p\cdot\varepsilon_\eta)e^{\theta_R\eta},
\]
the effect of this difference will be the existence of a rotation between dilaton and axion. 
The double copied amplitudes are obtained by applying the map \eqref{eq:MixedAmplitudes}, 
\[
\ampM_{\hel}&=-c_{++}\kappa\,m^2(u\cdot\varepsilon_\hel)^2\,e^{\bar{\theta}\eta}~,
\\
\ampM_{\bphi}&=\frac{\kappa\,c_{+-}}{2}\,p^2\,\cosh\Delta\theta=\frac{\charge\,m}{2}\,\,\cosh\Delta\theta~,
\\
\ampM_{\bB}&=\frac{\kappa\,c_{+-}}{2}\,p^2\,\sinh\Delta\theta=\frac{\charge\,m}{2}\,\,\sinh\Delta\theta~,\label{eq:MixedNUTAmplitudes}
\]
where we have defined $\charge=\kappa\,c_{+-}\,m$.
To test the effect of the rotation on the metric, let us compute the transformed Weyl tensor,
\[
\langle W^{\mu\nu\rho\sigma}(x) \rangle 
&=
 -\Re i \kappa^2 \,c_{++}\,m^2 \int \d\Phi(k) \, \del(2 k \cdot p)\, 
 e^{-i k \cdot x} \\
 &\hspace{1cm}
 \times\left[(\varepsilon_+\cdot u)^2  k^{[\mu} \varepsilon_-^{\nu]} k^{[\rho} \varepsilon_-^{\sigma]}\,e^{\bar\theta}
+ (\varepsilon_-\cdot u)^2 k^{[\mu} \varepsilon_+^{\nu]} k^{[\rho} \varepsilon_+^{\sigma]}\,e^{-\bar\theta} \right] \,.
\label{eq:explicitWeyl}
\]
A little algebra shows that this can be rewritten as 
\[
&\langle W^{\mu\nu\rho\sigma}(x) \rangle 
=
 -\Re i \kappa^2\,c_{++}\, m^2 \int \d\Phi(k) \, \del(2 k \cdot p)\, 
 e^{-i k \cdot x} \\
&\hspace{3cm}
\times\bigg[
\cosh\bar{\theta}
\left(
	k^{[\mu}u^{\nu]}k^{[\rho}u^{\sigma]}+\frac{1}{2}k^{[\mu}\eta^{\nu][\rho}k^{\sigma]}
\right)
\\
&\hspace{4cm}
-\frac{1}{2}\sinh\bar{\theta}\,\epsilon^{\mu\nu\tau\lambda}
\left(
	k_{[\tau}u_{\lambda]}k^{[\rho}u^{\sigma]}+\frac{1}{2}k_{[\tau}\delta_{\lambda]}^{[\rho}k^{\sigma]}
\right)
\bigg] \,.
\label{eq:finalWeyl}
\]
The first term, with the hyperbolic cosine, corresponds to the Schwarzschild solution, expression (3.26) of \cite{Monteiro:2020plf}, which we will denote $W^{\mu\nu\rho\sigma}_{\text{Schw.}}$. Making use of this notation leads to the compact result
\[
\langle W^{\mu\nu\rho\sigma} \rangle = \cosh\bar\theta\ W^{\mu\nu\rho\sigma}_{\text{Schw.}}-\sinh\bar\theta \,\frac{1}{2}\epsilon^{\mu\nu\tau \lambda}\ W_{\ \ \tau\lambda}^{\text{Schw.}\ \rho\sigma}.
\]
The second term represents the dual of $W_{\text{Schw.}}$, in analogy with the result of section \ref{sec:dyon}.
We conclude that the angle $\bar{\theta}$ indeed rotates the mass and the NUT charge of the solution \cite{Huang:2019cja,Alawadhi:2019urr}.

The Weyl tensor we have computed represents the graviton degrees of freedom in $\mathfrak{R}^{\mu\nu\rho\sigma}$. 
The next step is to obtain the classical expectation value of the dilaton and axion degrees of freedom. 
Instead of computing the corresponding components of $\mathfrak{R}^{\mu\nu\rho\sigma}$, we will obtain the field profiles $\langle \bphi\rangle$ and $\langle \bsigma\rangle$ directly.

Let us start with the classical expectation value of the dilaton field, $\langle\bphi\rangle=\langle\psi|S^\dagger\bphi S|\psi\rangle$. We use the field operator~\eqref{eq:DilatonOp}, exponentiation of the coherent state \eqref{eq:NSNSCoherent1} with the amplitude given in \eqref{eq:MixedNUTAmplitudes}. The result is
\[
\langle\bphi(x)\rangle=\charge\,m\,\cosh\Delta\theta\  \Re i\int \d\Phi(k)\, \del(2p\cdot k) e^{-ik\cdot x}~.
\]
Performing the integration as in \eqref{scalarkern}, we obtain
\[
\langle\bphi(x)\rangle
=
\charge\,\cosh\Delta\theta\ \frac{\Theta(\rho^2)}{8\pi}\,\frac{1}{\rho}
~.\label{eq:22NUTdilaton}
\]

We will now tackle the axion field. Recalling \eqref{eq:BfieldOp} and taking a derivative, we quickly find
\[
\langle \bH_{\mu\nu\rho}(x)\rangle
=\langle\frac{1}{2}\partial_{[\mu}\bB_{\nu\rho]}(x)\rangle
=\frac{\charge}{2}\,\sinh\Delta\theta\Re\int\d\Phi(k)\del(u\cdot k)k_{[\mu}\varepsilon_{\nu}^+\varepsilon^-_{\rho]}\,e^{-ik\cdot x}~.
\]
At this stage, it is very helpful to note that
\[
\epsilon^{\mu\nu\rho\sigma}k_{[\nu}\varepsilon^+_{\rho}\varepsilon^-_{\sigma]}
=
k^{[\mu}n^{\nu}\varepsilon_+^{\rho}\varepsilon_-^{\sigma]}k_{[\nu}\varepsilon^+_{\rho}\varepsilon^-_{\sigma]}
=-3!\,k^{\mu}~.
\]
Hence,
\[
\Rightarrow
\langle\epsilon^{\mu\nu\rho\sigma}H_{\nu\rho\sigma}(x)\rangle
&=-3\,\charge\,\sinh\Delta\theta\,\Re\int\d\Phi(k)\,\del(k\cdot u)\,k^\mu\,e^{-ik\cdot x}\\
&=-3\,\charge\,\sinh\Delta\theta\,\partial^{\mu}\left(\frac{\Theta(\rho^2)}{4\pi}\,\frac{1}{\rho}\right)~.
\]
This expression provides direct information on the axion $\sigma$. To see how, note from equation \eqref{eq:axiondef}, expanded to leading order,
that the relation between $H$ and $\sigma$ is simply
\[\label{leadax}
\bH_{\mu\nu\rho}=-\epsilon_{\mu\nu\rho\sigma}\,\partial^{\sigma}\bsigma
\Rightarrow \epsilon^{\mu\nu\rho\sigma}\bH_{\nu\rho\sigma}=-3!\,\partial^\mu\bsigma~.
\]
Comparing with the previous expression, we find
\[
\langle\bsigma(x)\rangle
=
\charge\,\sinh\Delta\theta\,\frac{\Theta(\rho^2)}{8\pi}\,\frac{1}{\rho}~.\label{eq:22NUTaxion}
\]
\subsection{Newman-Janis shift}
\label{sec:NJgravity}
Now we turn our attention to the spin parameter.
Just as we did in the previous section, we can use the prescription \eqref{eq:DCmap} to source an axion and a dilaton. However,  we now consider products of gauge theory amplitudes with different spins
\[\label{mappp}
\ampA^{(L)}_\eta=-2Q(p\cdot\varepsilon_\eta)e^{i\,\eta\, a_L\cdot k},
\\
\ampA^{(R)}_\eta=-2Q(p\cdot\varepsilon_\eta)e^{i\,\eta\, a_R\cdot k}.
\]
These yield the following gravity amplitudes 
\[
\ampM_{\hel}&=-\kappa\,c_{++}\,m^2(u\cdot\varepsilon_\hel)^2\,e^{i\, \eta\, \bar{a}\cdot k}~,
\\
\ampM_{\bphi}&=\frac{\charge\,m}{2}\,\,\cos (\Delta a\cdot k)~,
\\
\ampM_{\bB}&=\frac{\charge\,m}{2}\,\,\sin( \Delta a\cdot k)~.\label{eq:MixedKerrAmplitudes}
\]

Once more, the graviton components of the fat curvature tensor found in  \eqref{eq:lienarRiemannExpectation} reduce to the Weyl tensor
\begin{equation}
 W^{\mu\nu\rho\sigma}(x)
=\kappa\,c_{++}\,\Re\, i
\int \d\Phi(k)\del(2k\cdot p) e^{-ik\cdot x}
\sum_\eta\,\ampM_{\eta}\varepsilon_{-\eta}^{[\mu} k^{\nu]}\varepsilon_{-\eta}^{[\rho}k^{\sigma]} e^{i\eta k\cdot \bar a}.
\end{equation}

It is not difficult to see that this matches the classical computation with a spinning source.
The linearised EOMs are, writing ${g_{\mu\nu}(x)=\eta_{\mu\nu}+\kappa\, h_{\mu\nu}(x)}$
 \begin{equation}\label{efe}
 \partial^2 h^{\mu\nu}(x)=-\kappa \,{{\mathsf{P}}^{\mu\nu}}_{\alpha\beta} T^{\alpha\beta}(x), \qquad {{{\mathsf{P}}}^{\mu\nu}}_{\alpha\beta} =\frac{1}{4} {\delta^{(\mu}}_{(\alpha} {\delta^{\nu)}}_{\beta)}-\frac{1}{2}\eta^{\mu\nu}\eta_{\alpha\beta},
 \end{equation}
with the following stress-energy tensor for Kerr \cite{Guevara:2020xjx, justin2017}
\begin{equation}
T^{\mu\nu}(x)={m}\int \d\tau\, u^{(\mu} {\exp( \bar{a}*\partial)^{\nu)}}_\rho u^\rho \delta^{(4)}(x-u\tau) \,.
\end{equation}
Solving \eqref{efe} with the usual boundary conditions, we find  the linearised metric
\begin{equation}\label{hkerr}
\begin{split}
h^{\mu\nu}(x)&=-\kappa m^2 \text{Re}\,i \int{\d}\Phi(k)\hat{\delta}(p\cdot k)  e^{-ik\cdot x} \,
{{{\mathsf{P}}}^{\mu\nu}}_{\alpha\beta} \,u^{(\alpha} {\exp(-i \bar{a}*k)^{\beta)}}_\rho u^\rho
\\&=
-\kappa m^2 \text{Re}\,i \int{\d}\Phi(k)\hat{\delta}(p\cdot k)  e^{-ik\cdot x} \left[
\left(
u^\mu u^\nu -\frac{1}{2}\eta^{\mu\nu} 
\right)\cos( \bar{a}\cdot k)\right.  \\&
\qquad
\qquad
\qquad
\qquad
\qquad
\qquad
\qquad
\qquad
-\frac{i}{2}
u^{(\mu}\epsilon^{\nu)}(\bar{a}, k, u)\left.\frac{\sin( \bar{a}\cdot k)}{\bar{a}\cdot k}\right],
\end{split}
\end{equation}
from which the curvature can be computed. After some tedious but straightforward algebra one finds
\begin{equation} 
\begin{split}\label{WWWeyl}
W^{\mu\nu\rho\sigma}(x)&=
-\kappa^2 m^2 \text{Re}\,i  \sum_{\hel} \int{\d}\Phi(k)\hat{\delta}(2p\cdot k)  e^{-ik\cdot x} (\varepsilon_{\hel}\cdot u)^2 k^{[\mu}\varepsilon_{-\hel}^{\nu]} k^{[\rho}\varepsilon_{-\hel}^{\sigma]}e^{i\hel k\cdot \bar{a}}  .
\end{split}
\end{equation}
The result matches the one we obtained from amplitudes upon setting $c_{++}=1$.

For the dilaton and the axion, the calculations are formally analogous to the ones outlined in \ref{sec:TaubNUT}, except that now we have momentum dependent trigonometric functions which characterise the spin mixing. For this reason we omit the explicit computations and report the final results. We find for the dilaton 
\[
\langle\bphi(x)\rangle&=\frac{\charge}{2}\, \Re i\int \d\Phi(k)\, \del(u\cdot k) e^{-ik\cdot x} \cos (\Delta a\cdot k)~\\&=
\frac{\charge}{8}\left(
S_{\Delta a, 0}(x)+S_{-\Delta a, 0}(x)
\right),
\]
referencing to the definition of the scalar potential \eqref{scalarkern}.
The axion is instead given by
\[
\langle\epsilon^{\mu\nu\rho\sigma}\bH_{\nu\rho\sigma}(x)\rangle
&=-3\,\charge\,\partial^\mu \Re i\int\d\Phi(k)\,\del(k\cdot u)\,e^{-ik\cdot x} \sin (\Delta a\cdot k),
\]
telling us that the scalar $\bsigma$ 
is at leading order
\[
\langle\bsigma(x)\rangle&=\frac{\charge}{2} \Re i\int \d\Phi(k)\, \del(u\cdot k) e^{-ik\cdot x} \sin (\Delta a\cdot k)~\\&=
\frac{\charge}{8}\left(
S_{\Delta a, 0}(x)-S_{-\Delta a, 0}(x)
\right).
\]

\subsection{Comparison with known solutions}\label{Comparison with known solution}
The linearised solution obtained in section \ref{sec:TaubNUT} corresponds to an axi-dilaton Taub-NUT black hole. This solution is known exactly; see (17), (19) in \cite{Burgess:1994kq}. It is interesting to check that our results agree with the linearisation of the known solution.
There, dilaton and axion are given as\footnote{Ignoring factors of $\sqrt{3}$ that can be absorbed into $\delta$ at linear order..}
\[
e^{-\bphi}=(1+\epsilon^2)\frac{\Lambda^\delta}{\epsilon^2\Lambda^{2\delta}+1}~,\qquad
\bsigma=\frac{\epsilon(\Lambda^{2\delta}-1)}{\epsilon^2\Lambda^{2\delta}+1}~,
\]
where 
$$
\Lambda=1-\frac{R_0}{R}~,
$$
$\delta\,R_0$ is the charge of the dilaton and $\epsilon$ is a duality rotation parameter between dilaton and axion. Notice that $R$ is the (3,1) signature equivalent of $\rho$. At linearised level, the fields decouple and the metric is equivalent to Taub-NUT. Expanding at linear order the other fields and defining $\epsilon=-\tan\frac{\Delta\theta}{2}$, we find
\[
\bphi=
\cos\Delta\theta\,\frac{\delta\, R_0}{R}~,\qquad
\bsigma=\sin\Delta\theta\,\frac{\delta\,R_0}{R}~.
\]
Our solution \eqref{eq:22NUTdilaton}, \eqref{eq:22NUTaxion} agrees with this up to an overall constant ($\charge=16\pi\, \delta\, R_0$).\footnote{After rotating back to Minkowski signature, the hyperbolic trigonometric functions turn into standard trigonometric functions. Additionally, one has to continue $\sigma\to i\sigma$ due to its pseudo-scalar nature, which cancels the factor of $i$ from the sine.  The factor of $16$ takes into account a factor of $2$ generated by the analytic continuation of the propagator \eqref{eq:ContinuationScalar}.}
 Then,  $\Delta\theta$ is just the parameter inside $\text{SL}(2,\mathbb{R})$ that generates linear rotations between dilaton and axion.

In the special case where $\theta_L=\theta_R$ both single copies are identical, and we have no mixing: $\Delta\theta=0$. From \eqref{eq:MixedNUTAmplitudes}, we see that this implies that the axion will vanish, leaving a linearised solution that would be the equivalent to Taub-NUT plus the dilaton. In \cite{Burgess:1994kq}, this corresponds to (17) and (18).

On the contrary, if $\theta_R=-\theta_L$, $\bar{\theta}$ vanishes and the resulting metric has vanishing NUT charge. The result is a linearised Schwarzschild metric plus axion plus dilaton, corresponding to the linearisation of (10) and (13) in \cite{Burgess:1994kq}.

When both rotation angles are zero, both the NUT charge and axion vanish. We are left with a linearised JNW solution. This simplified solution will be used in section \ref{sec:positionSpace} to check for the existence of a double copy relation in position space. 

The solutions considered in section \ref{sec:NJgravity} involving spin are not so well understood in the literature. 
There have been attempts to apply a Newman-Janis shift to the JNW solution, with the prospects of obtaining a spinning generalisation. 
However, these claimed generalisations fail to satisfy the Einstein-dilaton equations of motion \cite{Bogush:2020lkp}. 
Although linear, our solution might help to find a satisfactory  generalisation of the JNW metric with spin.

\section{Heterotic double copy}
\label{sec:heterotic}

It is straightforward to get an extra set of gauge fields in the double copy, by simply supplementing one of the YM factors with a scalar $X$. For instance, let us assume that the `left gauge theory' is coupled to a bi-adjoint scalar\footnote{This should not be confused with the bi-adjoint scalar that appears in the denominator of various formulations of the double copy.}  which transforms under the symmetry groups $G$ and $\hat{G}$. The Lagrangians for our gauge theory factors are given by:
\begin{equation}
\label{YMandYMscalar}
\begin{aligned}
\mathcal{L}^{(L)}=&-\tfrac{1}{4}\text{Tr}\left[F_{\mu\nu}^{(L)}F^{(L)\mu\nu} \right]+\tfrac{1}{2}\text{Tr}\left[D_\mu X^{\hat{a}} D^\mu X_{\hat{a}} \right]-\tfrac{g^2}{4}\left(f_{abc}X^{b\hat{b}}X^{c\hat{c}}\right)\left(f^a_{\ b'c'}X^{b'}_{\ \hat{b}} X^{c'}_{\ \hat{c}} \right) \\
&+\tfrac{g g'}{3!}if_{abc}F_{\hat{a}\hat{b}\hat{c}}X^{a\hat{a}} X^{b\hat{b}} X^{c\hat{c}}  \,\,, \\
\mathcal{L}^{(R)}=&-\tfrac{1}{4}\text{Tr}\left[F_{\mu\nu}^{(R)}F^{(R)\mu\nu} \right]\,\,.
\end{aligned} 
\end{equation}
where $f_{abc}$ and $F_{\hat{a}\hat{b}\hat{c}}$ are the structure constants of $G$ and $\hat{G}$, respectively, and $g'$ is a dimensionful arbitrary constant. In the above, the trace is taken over the group $G$. We can think of $\mathcal{L}^{(L)}$ as coming from the dimensional reduction of a pure YM theory, together with the gauging of (a subset of) the global symmetry, which introduces the cubic interactions \cite{Chiodaroli:2014xia}. The double copy will then give an Einstein-Yang-Mills-dilaton-axion theory. Upon linearisation, the interaction terms in \eqref{YMandYMscalar} drop out.
For our purposes of studying the linearised solutions of heterotic gravity, it suffices to take our gauge groups to be $U(1)$, resulting in a single Maxwell field in the gravity theory. 
 
If we denote the amplitude for the 3-point interaction of the scalar $X$ with the static particle by $\ampA_X$, we can write schematically:
\begin{equation}
\left(\ampA_+^{(L)}+\ampA_-^{(L)} + \ampA_X\right)\ \otimes\ \left(\ampA_+^{(R)}+\ampA_-^{(R)}\right)
= \left\{\begin{matrix}
\vspace{7pt}
\ampA_\pm^{(L)}\otimes\ampA_\pm^{(R)}\to\ampM_\pm\\
\vspace{7pt}
\ampA_{(+}^{(L)}\otimes \ampA_{-)}^{(R)}\to\ampM_\phi\\ 
\vspace{7pt}
\ampA_{[+}^{(L)}\otimes \ampA_{-]}^{(R)}\to\ampM_B\\ 
\ampA_X\otimes\ampA_\pm^{(R)}\to \ampA_\pm
\end{matrix}\right. 
\end{equation}
where $\ampA_\pm$ is the 3-point amplitude for the gauge field in the heterotic theory. The amplitude $\ampA_X$ is a constant $\ampA_X=Q_X$. Then, the double copy for the graviton, dilaton and axion amplitudes, already given in \eqref{eq:MixedAmplitudes}, is supplemented by
\begin{equation}
\ampA_\pm=\frac{c_X}{Q_X}\ \ampA_X\ \ampA_\pm^{(R)}\,\,,\
\end{equation}
for some arbitrary constant $c_X$. The dilaton and axion fields are still given by \eqref{eq:22NUTdilaton} and \eqref{eq:22NUTaxion}, as they are unaffected by the addition of the gauge field in the linearised approximation. The gauge field will have electric and magnetic charges given by
\begin{equation}
\label{el_and_mag_ch_DC}
\begin{aligned}
Q_E^{d.c.}=&c_X\, Q \,\text{cosh}(\theta_R)\,\,,\\
Q_M^{d.c.}=&c_X\, Q\, \text{sinh}(\theta_R)\,\,.
\end{aligned} 
\end{equation}

\subsection{Comparison with known solution}
We now want to compare with \cite{Burgess:1994kq}. Here, the general spherically symmetric, static and asymptotically flat solution to the heterotic theory is obtained via a series of transformations. Starting with the graviton-dilaton solution, characterised by parameters $l$ and $\delta$, one can perform three $\text{SL}(2,\mathbb{R})$ transformations, with parameters $\omega,\epsilon$ and $\rho$\footnote{For each $\text{SL}(2,\mathbb{R})$, imposing the vanishing of the dilaton and the axion at infinity reduces the three parameters of $\text{SL}(2,\mathbb{R})$ to one.}, interspersed by a discrete T-duality and an $\text{O}(1,1)$ transformation with parameter $x$ to obtain the full family of axion-dilaton-Taub-NUT metric coupled to a vector field. The axion and dilaton charges are given by

\begin{equation}
\label{dil_grav_string}
\begin{aligned}
\hat{Q}_D&=\tfrac{1-\rho^2}{1+\rho^2}\bar{Q}_D-\tfrac{2\rho}{1+\rho^2}\bar{Q}_A\,\,,\\
\hat{Q}_A&=\tfrac{1-\rho^2}{1+\rho^2}\bar{Q}_A+\tfrac{2\rho}{1+\rho^2}\bar{Q}_D\,\,,
\end{aligned}
\end{equation}
with
\begin{equation}
\begin{aligned}
\bar{Q}_D&=\tfrac{l}{2}\left[(1+x)\tfrac{1-\epsilon^2}{1+\epsilon^2}\delta+(1-x)\tfrac{1-\omega^2}{1+\omega^2}\sqrt{1-\delta^2} \right]\,\,,\\
\bar{Q}_A&=\tfrac{l}{2}\left[(1+x)\tfrac{2\epsilon}{1+\epsilon^2}\delta+(1-x)\tfrac{2\omega}{1+\omega^2}\sqrt{1-\delta^2} \right]\,\,.
\end{aligned} 
\end{equation}
Using the notation $\rho=\tan\tfrac{\theta_\rho}{2}$, $\epsilon=\tan\tfrac{\theta_\epsilon}{2}$ and restricting to $\epsilon=\omega$, the charges can be shown to reduce to 
\begin{equation}
\begin{aligned}
\hat{Q}_D&=\text{cos}(\theta_\rho+\theta_\epsilon)\tfrac{l}{2}\left[(1+x)\delta+(1-x)\sqrt{1-\delta^2}  \right]\,\,,\\
\hat{Q}_A&=\text{sin}(\theta_\rho+\theta_\epsilon)\tfrac{l}{2}\left[(1+x)\delta+(1-x)\sqrt{1-\delta^2}  \right]\,\,.
\end{aligned} 
\end{equation}
Then, comparison with \eqref{eq:22NUTdilaton} and \eqref{eq:22NUTaxion} gives\footnote{As explained in \autoref{Comparison with known solution}, the hyperbolic trigonometric functions will turn into standard trigonometric functions upon rotating back to Minkowski signature.}
\begin{equation}
\begin{aligned}
\Delta\theta=&\theta_\rho+\theta_\epsilon\,\,,\\
\frac{\tilde{c}}{16\pi}=&\tfrac{l}{2}\left[(1+x)\delta+(1-x)\sqrt{1-\delta^2} \right]\,\,,
\end{aligned} 
\end{equation}
This reduces to the solution without the gauge field for $\theta_\rho=0$ and $x=1$ (note that $x=1$ corresponds to the identity matrix in the $\text{O}(1,1)$ transformation used in \cite{Burgess:1994kq}). The electric and magnetic charges in \cite{Burgess:1994kq}, with the restriction $\epsilon=\omega$, reduce to 
\begin{equation}
\label{el_and_mag_ch_strings}
\begin{aligned}
\hat{Q}_E=&\sqrt{x^2-1}l\left(\sqrt{1-\delta^2}-\delta\right) \cos\left(\tfrac{\theta_\rho}{2}+\theta_\epsilon\right)\,\,,\\
\hat{Q}_M=&\sqrt{x^2-1}l\left(\sqrt{1-\delta^2}-\delta\right) \sin\left(\tfrac{\theta_\rho}{2}+\theta_\epsilon\right)\,\,.
\end{aligned}
\end{equation}
Then, comparing with \eqref{el_and_mag_ch_DC}, we get the identifications
\begin{equation}
\begin{aligned}
\theta_R&=\tfrac{\theta_\rho}{2}+\theta_\epsilon\,\,,\\
 c_X\, Q&=\sqrt{x^2-1}l\left(\sqrt{1-\delta^2}-\delta\right) \,\,.
\end{aligned}
\end{equation}
Finally, we remark that the restriction we imposed on the parameters (i.e.~setting $\epsilon=\omega$) was only necessary in order to obtain a straightforward map. One can always write a completely general (though more cumbersome) map via
\begin{equation}
\begin{aligned}
\frac{\tilde{c}}{16\pi}&=&\sqrt{\hat{Q}_D^2+\hat{Q}_A^2}\,, \qquad\qquad
\Delta \theta&=&
\text{tan}^{-1} \left(\frac{\hat{Q}_A}{\hat{Q}_D} \right)\,\,,\\
c_XQ&=&\sqrt{\hat{Q}_E^2+\hat{Q}_M^2}\,,\qquad\qquad
\theta_R&=&\text{tan}^{-1}\left(\frac{\hat{Q}_M}{\hat{Q}_E}\right)\,\,.
\end{aligned} 
\end{equation}

\section{Double copy in position space} 
\label{sec:positionSpace}
We have found expressions that exhibit clear double copy relations in on-shell momentum space. In this section, we will discuss whether these straightforward double copy relations can be carried over to position space.
In particular, we will see how the position-space Weyl double copy relations \cite{Luna:2018dpt} emerge from amplitudes.
In most of this section, we will set $a_{L,R}=\theta_{L,R}=0$, since this simple scenario is enough to illustrate our points. To simplify the discussion, each term in the expectation value of \eqref{eq: linearised riemann operator fields} will be analysed individually. We will omit the axion term, since it vanishes when $a_{L,R}=\theta_{L,R}=0$. 

First, we shall consider the terms associated to the graviton amplitude, which we will denote by $\mathfrak{R}^{(h)}{}_{\mu\nu\rho\sigma}$. This is precisely the Schwarzschild Riemann (and Weyl) tensor considered in \cite{Monteiro:2020plf}. After some algebraic manipulations,
\[
\langle \mathfrak{R}^{(h)\,\mu\nu\rho\sigma}\rangle
=
-\Re\,i\,m^2\kappa^2
\int\d \Phi(k)\del(2k\cdot p)e^{-ik\cdot x}
\left(
	k^{[\mu}u^{\nu]} k^{[\rho}u^{\sigma]}
	+\frac{1}{2}k^{[\mu}\eta^{\nu][\rho}k^{\sigma]}
\right)~.
\]
The second term in the brackets makes the expression traceless. Alternatively, we can write
\[
\langle \mathfrak{R}^{(h)\,\mu\nu\rho\sigma}\rangle
=
-\mathcal{P}^{\mu\nu\rho\sigma}_{\tau\lambda\eta\omega}\,\Re\,i\,m^2\kappa^2
\int\d \Phi(k)\del(2k\cdot p)e^{-ik\cdot x}
	k^{[\tau}u^{\lambda]} k^{[\eta}u^{\omega]}
	~,
\]
where $\mathcal{P}^{\mu\nu\rho\sigma}_{\tau\lambda\eta\omega}$ projects out the trace, as in the definition of the Weyl tensor\footnote{This projector is sufficient for our purposes but more general possibilities are available.}
\[
W^{\mu\nu\nu\rho}&=\mathcal{P}^{\mu\nu\rho\sigma}_{\tau\lambda\eta\omega}\,R^{\tau\lambda\eta\omega}~,
\\
\mathcal{P}^{\mu\nu\rho\sigma}_{\tau\lambda\eta\omega}&=
\delta^{\mu}_{\tau}\delta^{\nu}_\lambda\delta^\rho_\eta\delta^\sigma_\omega
+\frac{1}{2}g_{\tau\eta}
\delta^{[\mu}_\lambda g^{\nu][\rho}\delta^{\sigma]}_\omega 
+\frac{1}{6}\,g_{\tau\eta}\,g_{\lambda\omega}\,g^{\mu[\rho}\,g^{\sigma]\nu}~.
\]
Next, we can take the factors of $k$ outside the integral as derivatives
\[
\langle \mathfrak{R}^{(h)\,\mu\nu\rho\sigma}\rangle
&=
\mathcal{P}^{\mu\nu\rho\sigma}_{\tau\lambda\eta\omega}\,\partial^{[\tau}u^{\lambda]} \partial^{[\eta}u^{\omega]}\,\Re\,i\,m^2\kappa^2
\int\d \Phi(k)\del(2k\cdot p)e^{-ik\cdot x}
\\
	&=
\mathcal{P}^{\mu\nu\rho\sigma}_{\tau\lambda\eta\omega}\,\partial^{[\tau}u^{\lambda]} \partial^{[\eta}u^{\omega]}\,\frac{\,m\kappa^2}{2}
\,\Re\,i
\int\d \Phi(k)\del(k\cdot u)e^{-ik\cdot x}
	~.
\]
For positive $t^1$, it can be checked that
\begin{equation}
\Re\,i\,\int\d \Phi(k)\del(k\cdot u)e^{-ik\cdot x}
=-\frac{1}{2}\int \dd^3k\, \frac{e^{-ik\cdot x}}{k^2}~.
\end{equation}
The integral on the right is performed over the three-dimensional subspace of momenta orthogonal to the worldline $k\cdot u=0$, so $k_\mu=(k_1,0,k_3,k_4)$. This prescription will be used in the rest of this section. 
Additionally, the divergence is resolved by the $i\epsilon$ prescription $(k)^2=(k^1+i\epsilon)^2-|\vec{k}|^2$, which selects the retarded contour.
Substituting in the Riemann tensor and taking the derivatives inside the integral, we get
\[
\langle \mathfrak{R}^{(h)\,\mu\nu\rho\sigma}\rangle
&=
\mathcal{P}^{\mu\nu\rho\sigma}_{\tau\lambda\eta\omega}
\frac{\,m\kappa^2}{4}
\int \dd^3k \,\frac{e^{-ik\cdot x}}{k^2} 
k^{[\tau}u^{\lambda]} k^{[\eta}u^{\omega]}~. \label{eq:intermediateRh}
\]
This already looks like a double copy of the field strength tensor
\begin{equation}
F^{\mu\nu}(x)=-Q\int\dd^3k\,\frac{e^{-ik\cdot x}}{k^2}k^{[\mu}u^{\nu]}~.
\end{equation}
To obtain a concrete double copy expression, we introduce a delta function
\[
\langle \mathfrak{R}^{(h)\,\mu\nu\rho\sigma}\rangle
&=
\mathcal{P}^{\mu\nu\rho\sigma}_{\tau\lambda\eta\omega}
\frac{\,m\kappa^2}{4}
\int \dd^3k\,\dd^3 q \,\del^3(k-q) \frac{e^{-ik\cdot x}}{k^2} 
k^{[\tau}u^{\lambda]} q^{[\eta}u^{\omega]}
\\
&=
\mathcal{P}^{\mu\nu\rho\sigma}_{\tau\lambda\eta\omega}
\frac{\,m\kappa^2}{4}
\int \dd^3k\,\dd^3 q\, \d^3 y\  e^{-iy\cdot(q-k)}\,\frac{e^{-ik\cdot x}}{k^2} 
k^{[\tau}u^{\lambda]} q^{[\eta}u^{\omega]}
\\
&=
-\mathcal{P}^{\mu\nu\rho\sigma}_{\tau\lambda\eta\omega}
\frac{\,m\kappa^2}{4}
\int \d^3 y\ \int \dd^3k\,\,\frac{e^{-ik\cdot (x-y)}}{k^2} 
k^{[\tau}u^{\lambda]}\,
\int\dd^3 q\, \  e^{-iy\cdot q}  q^{[\eta}u^{\omega]}
~.
\]
The last line is already a convolution of $F^{\tau\lambda}$ with the integral in $q$.
To complete the calculation we will need the scalar field and its formal inverse
\begin{equation}
S(x)=-\int\dd^3 k\,\frac{e^{-ik\cdot x}}{k^2}~,
\qquad 
S^{-1}(x)=-\int\dd^3 k\, e^{-ik\cdot x}\,k^2~,
\end{equation}
satisfying\footnote{The symbol $\conv$ denotes convolution: $(f\conv g)(x)=\int \d y f(y)g(x-y)$.} 
\[
\left(S\conv S^{-1}\right)(x)=\delta(t_1)\delta^2(\mathbf{x}) \,.
\]
We emphasise that the expression for $S^{-1}$ is only formal, due to the divergence of the integral. $S^{-1}$ is really defined operationally, acting via the convolution. It will always appear in convolutions where this divergence is cancelled, yielding a finite result.
 The calculation can be carried out following a strategy similar to the previous one. Inserting a delta function and a factor which equals $1$ on its support, we have
\[
\langle \mathfrak{R}^{(h)\,\mu\nu\rho\sigma}\rangle
&=
-\mathcal{P}^{\mu\nu\rho\sigma}_{\tau\lambda\eta\omega}
\frac{\,m\kappa^2}{4}\ \int \d^3 y\ 
\int \dd^3k\,\,\frac{e^{-ik\cdot (x-y)}}{k^2} k^{[\tau}u^{\lambda]}\\
&\hspace{5cm}\int \,\dd^3 q\,\dd^3l\,\del(l-q)\,  \frac{l^2}{q^2} \,e^{-iy\cdot q} q^{[\eta}u^{\omega]}
\\
&=
-\mathcal{P}^{\mu\nu\rho\sigma}_{\tau\lambda\eta\omega}
\frac{\,m\kappa^2}{4}
\int \d^3y\,\,\d^3z \,
\int \dd^3k\,\,\frac{e^{-ik\cdot (x-y)}}{k^2} k^{[\tau}u^{\lambda]}\\
&\hspace{5cm}\int\dd^3 q\,\frac{1}{q^2} \,e^{-iq\cdot(y-z)} q^{[\eta}u^{\omega]}\,\int\dd^3l\,l^2\,e^{-il\cdot z}\,  
~.
\]
Finally, we recognise another convolution,
\[
\langle \mathfrak{R}^{(h)\,\mu\nu\rho\sigma}(x)\rangle
&=
-\frac{\,m\,\kappa^2}{4\,Q^2}\mathcal{P}^{\mu\nu\rho\sigma}_{\tau\lambda\eta\omega}\ \left(F^{\tau\lambda} \conv S^{-1}\conv F^{\eta\omega}\right)(x)~.
\]
We conclude that this contribution of the Riemann tensor is the convolution of two copies of the field strength tensor with an inverse power of the scalar field.  We remark that the convolutions are performed over a 3-dimensional subspace of spacetime, reflecting the fact that all our solutions are independent of $t_2$.

The convolution we have obtained is the most natural operation from the point of view of the double copy \cite{Anastasiou:2014qba,LopesCardoso:2018xes,Anastasiou:2018rdx,Borsten:2019prq,Borsten:2020xbt,Borsten:2021zir}. 
However, we know that for some cases (like Schwarzschild) the relation must factorise in position space, turning convolutions into ordinary products. 
In this factorisation, the projector plays an important role. On its own, $F\conv S^{-1}\conv F$ does not factorise, but the offending terms are pure traces that are projected out by $\mathcal{P}$, leaving a neat factorised expression. In order to gain a better understanding of this, let us go back to \eqref{eq:intermediateRh} and take the derivatives out:
\[
\langle \mathfrak{R}^{(h)\,\mu\nu\rho\sigma}\rangle
&=
-\mathcal{P}^{\mu\nu\rho\sigma}_{\tau\lambda\eta\omega}\partial^{[\tau}u^{\lambda]} \partial^{[\eta}u^{\omega]}
\frac{\,m\kappa^2}{4}
\int \dd^3k \frac{e^{-ik\cdot x}}{k^2} 
\\
&=
\frac{\,m\kappa^2}{4}\mathcal{P}^{\mu\nu\rho\sigma}_{\tau\lambda\eta\omega}\partial^{[\tau}u^{\lambda]} \partial^{[\eta}u^{\omega]}
\,S(x)~.\label{eq:PddS}
\]
The crucial point now is that the double derivative action on $S(x)$ factorises into single derivatives under the contraction of the projector $\mathcal{P}$, in the sense to be described below. This happens in the strict interior of the split signature light-cone, where the curvature is non-vanishing and type D.
It is simpler if we first perform an analytic continuation to (1,3) signature to prove the factorisation using the properties of Kerr-Schild vectors recently reviewed in \cite{Luna:2016due}. First, we note that the analytic continuation of the scalar $S(x)$ is 
\begin{equation}
S(x)=\frac{1}{4\pi R}~,
\end{equation}
where $R$ can be interpreted as the retarded null distance between a point $x^\mu$ and a static worldline $y^\mu(\tau)$ tangent to $u^\mu$. Similarly, we can define a Kerr-Schild vector
\[
K^\mu=\frac{[x^\mu-y^\mu(\tau)]_{\rm ret}}{R}~.
\]
It is not hard to prove 
\begin{equation}
\partial_\mu R=u_\mu-K_\mu
~,\qquad 
\partial_\mu K_\nu=\frac{1}{R}\left(\eta_{\mu\nu}+K_\mu K_\nu-K_{\mu}u_{\nu}-K_{\nu}u_{\mu}\right)~.
\end{equation}
These two identities imply
\[
\partial_\mu S&=-4\pi\,S^2\,(u_\mu-K_\mu)
\\
\partial_\mu\partial_\nu S
&=
 3(4\pi)^2\,S^3\,(u_\mu-K_\mu)( u_\nu-K_\nu)
 +(4\pi)^2S^3\left(\eta_{\mu\nu}-u_\mu u_\nu\right)
~.
\]
The last line can be rewritten as
\[
\partial_\mu\partial_\nu S
=
3\,\frac{\partial_\mu S\,\partial_\nu S}{S}
+(4\pi)^2\,S^3\left(\eta_{\mu\nu}- u_\mu u_\nu\right)\label{eq:ddS}
~.
\]
Upon substitution in \eqref{eq:PddS}, the contribution from the last term on the right-hand side of the expression above vanishes. Hence,
\[
\mathcal{P}^{\mu\nu\rho\sigma}_{\tau\lambda\eta\omega}\partial^{[\tau}u^{\lambda]} \partial^{[\eta}u^{\omega]}
\,S(x)
=\frac{3}{S(x)}\mathcal{P}^{\mu\nu\rho\sigma}_{\tau\lambda\eta\omega}\left(\partial^{[\tau}u^{\lambda]}S(x)\right)\left(\partial^{[\eta}u^{\omega]}S(x)\right)~.
\label{eq:ScalarFactorisation}
\]
This expression completes the argument, because the factors in parenthesis equal the field strength tensor up to some constants,
\[
\langle \mathfrak{R}^{(h)\,\mu\nu\rho\sigma}\rangle
&=\frac{3\kappa^2\,m}{4}\,\mathcal{P}^{\mu\nu\rho\sigma}_{\tau\lambda\eta\omega}\,\frac{F^{\tau\lambda}F^{\eta\omega}}{S}
~.\]
This is the Weyl double copy relation for Schwarzschild, which is also valid in higher dimensions for an appropriate constant factor.

Above, we have presented the proof for the simplest example, where the deformation parameters $\bar{\theta}$ and $\bar{a}$ are set to zero. 
Since the Kerr-Taub-NUT solution satisfies the Weyl double copy, a similar factorisation must happen for generic $\bar{\theta}$ and $\bar{a}$. 
The proof is rather straightforward. 
The key observation is that \eqref{eq:ScalarFactorisation} holds also for $S_{\bar{a},\bar{\theta}}$, since the effects of $\bar{a}$ and $\bar{\theta}$ are a translation and a constant scaling respectively. 
Then, after some algebra, one can prove that
\[
\langle \mathfrak{R}^{(h)\,\mu\nu\rho\sigma}\rangle
&=\frac{3\kappa^2\,m}{4}\mathcal{P}^{\mu\nu\rho\sigma}_{\tau\lambda\eta\omega}\left(
\frac{F_-^{\tau\lambda}F_-^{\eta\omega}}{S_{\bar{a},\bar{\theta}}}+
\frac{F_+^{\tau\lambda}F_+^{\eta\omega}}{S_{-\bar{a},-\bar{\theta}}}
\right)
~,\]
where $F_+$ and $F_-$ are the self-dual and anti-self-dual parts of the field strength as defined in \eqref{eq:Fplusmindef}. This result represents the Weyl double copy expressed in tensorial form, a map that was also studied in \cite{Alawadhi:2019urr, Alawadhi:2020jrv}.

It would be interesting to see whether these simple position-space double copy relations can be extended to include the dilaton. We start from the linear dilaton contribution in position space,
\[
\langle \mathfrak{R}^{(\bphi)\,\mu\nu\rho\sigma}(x)\rangle
=
\frac{\charge\,m\,\kappa}{2}\,\Re\,i\,
\int\d \Phi(k)\del(2k\cdot p)e^{-ik\cdot x}
k^{[\mu}\eta^{\nu][\rho}k^{\sigma]}
~.
\]
On the support of the delta functions, we can write this as
\[
\langle \mathfrak{R}^{(\bphi)\,\mu\nu\rho\sigma}(x)\rangle
=
-\frac{\charge\,m\,\kappa}{2}\,\Re\,i
\int\d \Phi(k)\del(2k\cdot p)e^{-ik\cdot x}
f^{\lambda[\mu}\eta^{\nu][\rho}f^{\sigma]}{}_{\lambda}
~,\label{eq:DilatonRiemann}
\]
where $f^{\mu\nu}=k^{[\mu}u^{\nu]}$.
We can follow the same steps as before to obtain a position space convolutional double copy
\[
\langle \mathfrak{R}^{(\bphi)\,\mu\nu}{}_{\rho\sigma}(x)\rangle
=
-\frac{\charge\,\kappa}{8\,Q^2}\left( F^{\lambda[\mu}\conv S^{-1}\conv F_{\lambda[\rho}\right)\!(x)\ \,{\delta^{\nu]}}_{\sigma]}
~.
\]
In contrast to what happened for the graviton contribution, these convolutions cannot be turned into products. To see that, we could proceed as for the graviton and rewrite \eqref{eq:DilatonRiemann} as a second order differential operator acting on the Lorentz continuation of  $S(x)$. Then, we would like to use \eqref{eq:ddS} to accomplish the factorisation. The result would be
\[
\langle \mathfrak{R}^{(\bphi)\,\mu\nu}{}_{\rho\sigma}(x)\rangle
=\frac{\charge\,\kappa}{8}
\left(
	3\frac{F^{\lambda[\mu}\,\delta^{\nu]}_{[\rho}\,F_{\sigma]\lambda}}{S}
	+(4\pi)^2 S^3
	\left(
		\delta^{[\mu}_{[\rho}\delta^{\nu]}_{\sigma]}-
		\delta^{[\mu}_{[\rho}u^{\nu]}\,u_{\sigma]}
	\right)
\right)~.
\]
While the first term exhibits a local position-space double copy form, the others do not. Thus, the double copy of the dilatonic contribution is natural only in terms of convolutions and it is non-local in position space. Interestingly, the JNW solution admits an exact double copy interpretation in position space, based on a Kerr-Schild-like construction in double field theory \cite{Kim:2019jwm}. However, unlike the Kerr-Schild double copy for Schwarzschild, the dilatonic deformation makes the relation non-local in position space.

\section{Discussion}

Our understanding of the link from scattering amplitudes to classical solutions has matured. It has become straightforward 
to construct the linearised solutions associated to massive three-point amplitudes in four dimensions.
The technical developments which have clarified this link are the KMOC formalism~\cite{Kosower:2018adc,Maybee:2019jus,Cristofoli:2021vyo}, which constructs classical observables from amplitudes,
and the analytic continuation~\cite{Monteiro:2020plf} to split signature metrics. 
This second step allows us to use the methods of KMOC in the context of three-point amplitudes.

In this paper, we took advantage of these developments to perform a comprehensive survey of the classical fields
constructed by the double copy from gauge theory three-point amplitudes. 
As anticipated by previous work on the classical double copy~\cite{Monteiro:2014cda,Luna:2015paa,Luna:2018dpt,Huang:2019cja,Emond:2020lwi},
we saw that magnetic charge in gauge theory indeed double copies to NUT charge in gravity.
Furthermore, our methods confirmed the three-point amplitudes associated
by more indirect arguments~\cite{Huang:2019cja,Emond:2020lwi} to Taub-NUT and its spinning generalisation, Kerr-Taub-NUT.
Indeed a split-signature form of Kerr-Taub-NUT and its scattering amplitude also appeared very recently in independent work~\cite{Crawley:2021auj,Guevara:2021yud}.

The most basic example of a classical double copy is that from Coulomb to Schwarzschild~\cite{Monteiro:2014cda}.
However, the literature also contains a \emph{different} double copy of Coulomb: namely the JNW solution~\cite{Goldberger:2016iau,Luna:2016hge}. The origin of this non-uniqueness was discussed in \cite{Kim:2019jwm}, where a Kerr-Schild-type exact double copy interpretation of JNW was also presented, and in \cite{Luna:2020adi}, where an off-shell convolutional approach  based on the BRST formulation (including Fadeev-Popov ghosts) \cite{Anastasiou:2018rdx} was used. We are also able to understand the origin of the non-uniqueness using the framework introduced in our paper. It arises directly from choices inherent in the standard double copy of scattering amplitudes. At the level of amplitudes, it is always possible to \emph{define} the gravitational theory by declaring that
its three-point amplitudes are either the double copy of two same-helicity gluons (resulting in Einstein gravity) or the double copy of two
same-helicity gluons \emph{and} two opposite-helicity gluons (resulting in NS-NS gravity). 
Making the former choice, the double copy of Coulomb is indeed Schwarzschild.
The latter choice, by contrast, leads to the JNW solution. So we are free to choose the couplings of the massive particle.

It is fascinating that the Kerr solution corresponds to a particularly simple three-point amplitude~\cite{Arkani-Hamed:2019ymq}. 
Clearly, this fact is related to the Newman-Janis shift, which is an all-orders property of Kerr~\cite{Newman:1965tw}.
The ``single'' copy of the Kerr solution, $\sqrt{\text{Kerr}}$, is a solution of the Maxwell equations which is also endowed with a simple three-point
amplitude. Turning on a magnetic charge in addition to the spin leads to a spinning dyonic solution which is, to date, the most
general known three-point amplitude in pure gauge theory. 
The double copy of this amplitude in pure gravity is the Kerr-Taub-NUT solution~\cite{Emond:2020lwi}. 
However, it is also possible to perform the double copy of these amplitudes in NS-NS gravity where, as we have seen,
the resulting class of solutions is of the type Kerr-Taub-NUT-dilaton-axion. 
This generalises the previous discussion of the double copy from Coulomb to the JNW solution to the more general three-point amplitudes.

The double copy of gauge theory contains even more general possibilities. 
Including scalar matter in the definition of the gauge theory introduces a new three-point amplitude: this time between
the pointlike source and the scalar field.
The resulting ``heterotic'' double copy is a gravitational theory including dilatons and axions as well as gauge fields.
To date, this is the most general class of theories in which we can explicitly relate the double copy for three-point amplitudes
to explicit classical fields.

Our work also dealt with another apparent mystery in the classical double copy. For scattering amplitudes, the double copy
is clearly a creature of momentum space: it is local in that context. 
Yet the classical double copy is frequently presented in position space --- the question is then how does locality in
position space somehow become locality in momentum space? 
In section~\ref{sec:positionSpace}, we showed that this locality arises non-trivially only in specific cases which include the Kerr-Taub-NUT
solution.

In spite of this progress, there is still much to understand. 
The classical double copy obviously applied to solutions which are not (yet) connected to scattering amplitudes. 
An important example is the (A)dS-Schwarzschild metric, which is related by the Kerr-Schild double copy to an electromagnetic
solution with a point charge immersed in a background of constant charge density~\cite{Luna:2015paa}. 
Given that the classical double copy connects to scattering amplitudes as well as configurations with a cosmological constant,
perhaps it can lead to some insight on the double copy in the presence of a cosmological constant.

We focussed throughout on four-dimensional metrics. 
Of course this case is particularly important, and (since gravitational wave phenomenology motivates much of the recent interest in amplitudes and classical gravity) it is also true that particular attention has been given to the relevant amplitudes
in four dimensions.
However, black holes in higher dimensions have a number of fascinating properties which have received intense scrutiny in
the literature. 
It would be interesting to generalise our methods to this case. 
The spinorial structure of the generalised curvature which was so helpful for us would need to be understood; we
expect that this should link to the higher-dimensional spinorial decomposition of the curvature described in reference~\cite{Monteiro:2018xev}.

Finally, let us highlight an outstanding mystery facing the classical double copy.
In this article, we focused on linearised solutions of gravity. 
But in terms of the single copy, our solutions are exact. 
In the case of Schwarzschild (and indeed Kerr-Taub-NUT), the exact Weyl tensor, in an appropriate tetrad, is linear in mass (and NUT charge) and so equals its linearised approximation.
However, this is not the case in general.
How should we understand the curvature corrections in these cases from the perspective of amplitudes? 
Perhaps we can turn to the beautiful computations of Mougiakakos and Vanhove~\cite{Mougiakakos:2020laz} for further insight.
A related, but perhaps more difficult, question is what is the meaning of the Kerr-Schild property of certain spacetime metrics in terms
of scattering amplitudes?

\section*{Acknowledgements}

We are grateful to Rashid Alawadhi, Kanghoon Lee, Nathan Moynihan and Chris White for related discussions. SN is supported by STFC grant ST/T000686/1. RM and DPV are supported by the Royal Society via a University Research Fellowship and a Studentship Grant, respectively.

\appendix
\section{Conventions}\label{sec:Conventions}
We work in a Minkowski spacetime with signature 
\begin{equation}
\eta_{\mu\nu}=\text{diag}(+1, -1, -1, -1),
\end{equation}
which is  then continued to the split signature  one 
\begin{equation} 
\eta_{\mu\nu}=\text{diag}(+1, +1, -1, -1).
\end{equation}
Our Fourier transforms conventions are 
\[
f(x) &= \int \dd^4 k \, e^{-i k \cdot x} \, \tilde f(k) \,,\\
\tilde f(k) &= \int \d^4 x \, e^{ik\cdot x} \, f(x) \,,
\]
where to clean up factors of $2\pi$ we write
\[
\dd^nk =\frac{\d^n k }{(2\pi)^n}\,,\qquad \del^n(k)=(2\pi)^n \delta^n(k) \,.
\]
We also define (anti-)symmetrized brackets as
\[
v^{(\mu}w^{\nu)}=v^\mu w^\nu+v^\nu w^\mu, \qquad v^{[\mu}w^{\nu]}=v^\mu w^\nu-v^\nu w^\mu.
\label{eq:antisymmDef}
\]

\section{Volume form and null tetrad}\label{sec:VolumeForm}
This appendix is devoted to relating the null tetrad to the volume form. In four-dimensional spacetimes, any four-form must be proportional to the volume form. In particular,

\begin{equation}
k_{[\mu}n_{\nu}\varepsilon_{+\,\rho}\varepsilon_{-\,\sigma]}
\propto
\epsilon_{\mu\nu\rho\sigma}~.
\end{equation}
The proportionality factor can be obtained as follows. First, note that
\begin{align}
\epsilon^{\mu\nu\rho\sigma}\epsilon_{\mu\nu\rho\sigma}=4!~.
\end{align}
Next, we square the antisymmetrised tetrad,
\begin{align}
k_{[\mu}n_{\nu}\varepsilon_{+\,\rho}\varepsilon_{-\,\sigma]}k^{[\mu}n^{\nu}\varepsilon_{+}{}^{\rho}\varepsilon_{-}{}^{\sigma]}=4!(k\cdot n)^2(\varepsilon_+\cdot\varepsilon_-)^2=4!~.
\end{align}
Comparing both equalities, we conclude that 
\begin{equation}
\epsilon_{\mu\nu\rho\sigma}=\pm k_{[\mu}n_{\nu}\varepsilon_{+\,\rho}\varepsilon_{-\,\sigma]}~.
\end{equation}
The sign on the right is not fixed, as it depends on the particular choice of tetrad. Right-handed tetrads satisfy the relation with the plus sign. Left-handed tetrads can be turned into right-handed ones by interchanging $\varepsilon_+$ and $\varepsilon_-$. Therefore, without loss of generality, we can assume that the tetrad is right-handed.

Finally, we can obtain an useful relation by contracting the volume form with $k_\rho$ and $u_\sigma$,
\[
\epsilon^{\mu\nu\rho\sigma}k_{\rho}\,u_\sigma
&=
-\varepsilon_-\cdot u\,\epsilon^{\mu\nu\rho\sigma}k_\rho\varepsilon_{+\,\sigma}-\varepsilon_+\cdot u\,\epsilon^{\mu\nu\rho\sigma}k_\rho\varepsilon_{-\,\sigma}
\\
&=-k^{[\mu}n^{\nu}\varepsilon^{+\,\rho}\varepsilon^{-\,\sigma]}
\left(
	\varepsilon_-\cdot u\,k_\rho\varepsilon_{+\,\sigma}
	+
	\varepsilon_+\cdot u\,k_\rho\varepsilon_{-\,\sigma}
\right)
\\
&=-
\left(	
	\varepsilon_-\cdot u\, k^{[\mu}\varepsilon_+^{\nu]}
	-
	\varepsilon_+\cdot u\, k^{[\mu}\varepsilon_-^{\nu]}
\right)\label{eq:epsku}~.
\]

\section{Gravitational coherent state}\label{sec:CoherentState}
This section describes the exponentiation of the coherent state in the presence of the graviton, dilaton and axion fields.
Given that our initial state contains only one creation operator for the massive particle, 
the S-matrix can be expanded in multiplicities as
\[
S \ket{\psi}= \frac{1}{\mathcal{N}} (1 + i T_3 + i T_4 + \cdots) \ket{\psi} \,,
\]
where the $T_n$ are defined by
\[
T_{n+2} &= \frac{1}{n!} \sum_{\hel_1, \ldots, \hel_n} \int  \d\Phi(p') \d\Phi(p) \prod_{i=1}^n \d\Phi(k_i) \, 
\sum_{m=0}^n
\frac{n!}{l!\,m!\,(n-l-m)!}
\\&
\ampM^{(l,m,n-m-l)}_{-\hel_1, \ldots, -\hel_m}  (p \rightarrow p', k_1, \cdots, k_n) 
\times \del^4\left(p-p' - \sum k_i\right) \, a^\dagger(p') a(p)
\\
&\qquad\qquad a^\dagger_{\hel_1}(k_1) \cdots a^\dagger_{\hel_l}(k_m)a^\dagger_{\phi}(k_{l+1})\cdots a^\dagger_{\phi} (k_{l+m})\,a^\dagger_b(k_{l+m+1})\dots a^\dagger_B(k_{n})  \,.
\label{eq:Tn+2def}
\]
$\ampM^{(l,m,n-m-l)}_{-\hel_1, \ldots, -\hel_m} (p \rightarrow p', k_1 \cdots k_n) $ denotes the amplitude with two external scalar legs, $l$ graviton external legs, $m$ dilaton legs and $n-l-m$ axion legs. 
\\
The simplifications of the amplitude in the classical limit are equivalent to those exposed in \cite{Monteiro:2020plf}. First of all, loops are subleading in $\hbar$, so we should consider tree amplitudes exclusively. Then, all the higher multiplicity vertices and graviton-dilaton/axion vertices yield higher order terms in $\hbar$. Thus, only the 3-point vertices with two massive scalar legs
contribute. Finally, the fact that gravitons, dilatons and axions have momenta of order $\hbar$ implies that we can ``cut" the three-point amplitudes, simplifying the leading classical contribution to
\[
i \ampM&^{(l,m,n-l-m)} 
= 
\left(
	\prod_{i=1}^l i \ampM_{-\hel_i}(k_i) 
	\prod_{i=l+1}^{l+m} i \ampM_{\phi}(k_i)
	\prod_{i=l+m+1}^{n} i \ampM_{B}(k_i)
\right)\times
\\
&
\sum_{\pi} \frac{i}{2 p\cdot k_{\pi(1)} + i \epsilon} \frac{i}{2 p\cdot (k_{\pi(1)} + k_{\pi(2)}) + i \epsilon} \cdots 
\frac{i}{2 p\cdot (k_{\pi(1)} + k_{\pi(2)} + \cdots k_{\pi(n-1)}) + i \epsilon} \,.
\label{eq:n+2step}
\]
The second line will be simplified together with the total momentum conservation delta function thanks to 
\[
\del\left(\sum_{i=1}^n \omega _i\right) \sum_{\pi} \frac{i}{\omega_{\pi(1)} + i \epsilon} \frac{i}{\omega_{\pi(1)} + \omega_{\pi(2)} + i \epsilon} \cdots \frac{i}{\omega_{\pi(1)} + \omega_{\pi(2)} + \cdots \omega_{\pi(n-1)} + i \epsilon} \\
= \del(\omega_1) \del(\omega_2) \cdots \del(\omega_n) \,.
\]
Finally, we notice that the factors arrange themselves into a power of the sum of amplitudes
\[
iT_{n+2} \ket{\psi} &= \frac{1}{n!} \int \d\Phi(p) \,\varphi(p) \bigg(
	\sum_\hel \int \d \Phi(k)\, \del(2 p \cdot k) 
	\,i\ampM_{-\hel}(k)\, a^\dagger_\hel(k) 
	\\
	&\hspace{110pt}
	+\int d\Phi(k)\d \Phi(k)\, \del(2 p \cdot k) \,i\ampM_{\phi}(k)\, a^\dagger_\phi(k) \\
	&\hspace{120pt}
	+\int d\Phi(k)\d \Phi(k)\, \del(2 p \cdot k) \,i\ampM_{B}(k)\, a^\dagger_B(k) 
\bigg)^n \ket{p} \,.
\]
Consequently, summing over $n$ reproduces the expansion of an exponential, yielding the coherent state
\[
\hspace{-4pt}
S \ket{\psi} 
&=
 \frac{1}{\mathcal{N}} \int \d\Phi(p) \varphi(p)
\exp \bigg[ 
	 \int \d \Phi(k)\,  
	\del(2 p \cdot k)\, i 
	\\
	&\hspace{40pt}\left(
		\sum_\hel \ampM_{-\hel}(k) \,a^\dagger_\hel(k) 
		+\ampM_{\phi}(k) \,a^\dagger_\phi(k) 
		+\ampM_{B}(k) \,a^\dagger_B(k) 
	\right)
\bigg] \ket{p} \,.
\label{eq:NSNSCoherent2}
\]

\section{2-Spinors in Riemann-Cartan geometries}
\label{sec:spinorsRC}
In this appendix, we will study the Riemann-Cartan objects defined in section \ref{sec:generalisedCurvature} under the light of the 2-spinor formalism.
The generalisation of spinors to spacetimes with torsion was also addressed in \cite{Penrose:1983mf,Penrose:1985bww}.

The procedure to define the spinorial structure does not differ from the Riemannian case. 
Tensors are mapped to spinors using the Pauli matrices (also called Infeld-van der Waerden symbols) ${\sigma_\mu}^{A\Ad}$ and ${\tilde{\sigma}_\mu}^{\,\,\,\dot A A}$. 
The metric on spinor space is the anti-symmetric two by two matrix $\epsilon_{AB}$ (and $\depsilon_{\Ad \Bd}$). The conventions for raising and lowering spinors are
\begin{equation}
\xi^A=\epsilon^{AB}\xi_B~,\qquad \xi_A=\xi^{B}\epsilon_{BA}~,
\end{equation}
\begin{equation}
\epsilon^{AC}\epsilon_{CB}=\epsilon^A{}_B={\delta^A}_B~,\qquad
\epsilon^{CA}\epsilon_{CB}=\epsilon_{B}{}^{A}=-{\delta^A}_B~.
\end{equation} 
Similar expressions hold for $\depsilon_{\Ad\Bd}$. 
For conciseness, the $\sigma$-matrices will be used implicitly every time indices are translated from the spacetime tangent bundle to the spinorial bundles. In this way, we write
\begin{equation}
K_{\mu\nu\rho} \to K_{A\Ad B\Bd C\Cd}~.
\end{equation}
In going to spinorial space, we lengthen the list of indices, but we gain extra simplification power. 
This is because tensorial symmetries imply that the spinorial counterparts must decompose into lower rank symmetric spinors and epsilon matrices. 
Moreover, since every pair of indices is the sum of their symmetrisation plus their antisymmetrisation, any spinor can be expressed as a sum of totally symmetric spinors combined with epsilon matrices. 
The most famous example is the reduction of the Riemann spinor to its irreducible components
\begin{equation}
\begin{aligned}
R_{A\Ad B\Bd C\Cd D\Dd}
&=
\Psi_{ABCD}\depsilon_{\Ad\Bd}\depsilon_{\Cd\Dd}
+\tilde\Psi_{\Ad\Bd\Cd\Dd}\epsilon_{AB}\epsilon_{CD}
\\
&\hspace{1cm}
+\Phi_{AB\Cd\Dd}\depsilon_{\Ad\Bd}\epsilon_{CD}
+\tilde\Phi_{\Ad\Bd CD}\epsilon_{AB}\depsilon_{\Cd\Dd}
\\
&\hspace{2cm}
+2\Lambda\,(\epsilon_{AC}\,\epsilon_{BD}\,\depsilon_{\Ad\Cd}\,\depsilon_{\Bd\Dd}-\epsilon_{AD}\,\epsilon_{BC}\,\depsilon_{\Ad\Dd}\,\depsilon_{\Bd\Cd})~.
\end{aligned}
\end{equation}
The symmetry under the exchange of pairs of indices and the first Bianchi identity impose $\tilde\Phi_{\Ad\Bd CD}=\Phi_{CD\Ad\Bd}$ and $\tilde\Lambda=\Lambda$ respectively.

However, this result does not hold for Riemann-Cartan manifolds. One of our goals is to see explicitly how the above expression changes in the presence of contorsion. 
As a preliminary step, we have to study the contorsion itself from the point of view of spinors.

\subsection{Contorsion spinors}
 The natural first step for decomposing the contorsion spinor is to  exploit the antisymmetry of $K_{\mu\nu\rho}$ in the first and third indices,
\begin{equation}
K_{A\Ad B\Bd C\Cd}=\Theta_{ABC\Bd}\depsilon_{\Ad\Cd}+\tilde\Theta_{\Ad\Bd\Cd B}\epsilon_{AC}~,
\end{equation}
where $\Theta_{ABC\Bd}=\frac{1}{2}\Theta_{(A|B|C)\Bd}$.  The resulting spinor is still not totally symmetric, implying that it can be separated into two irreducible parts
\begin{equation}
\Theta_{ABC\Bd}=
\Xi_{C\Bd}\,\epsilon_{AB}
+\Xi_{A\Bd}\,\epsilon_{CB}
+\Omega_{ABC\Bd}~,
\end{equation}
where $\Omega_{ ABC\Bd}=\frac{1}{3!}\Omega_{( ABC)\Bd}$.  The spinors $\Xi_{A\Bd}$ and $\Omega_{ ABC\Bd}$ constitute  the irreducible spinorial decomposition of the contorsion.

Now that we have pinned down the spinorial degrees of freedom of the contorsion, we can map them to the tensorial degrees of freedom. 
These tensorial degrees of freedom are arranged into three components: a completely antisymmetric tensor $\breve{K}_{\mu\nu\rho}$, a trace $\bar{K}_\mu$ and a traceless tensor $\hat{K}_{\mu\nu\rho}$
\begin{subequations}\label{eq:ContorsionTensorComponents}
\begin{align}
\breve{K}_{\mu\nu\rho}&=\frac{1}{3!}\,K_{[\mu\nu\rho]}~,\\[0.5em]
\bar{K}_\mu\ \ &=K^\nu{}_{\nu\mu}~,\\
\hat{K}_{\mu\nu\rho}&=\frac{1}{3}(K_{(\mu\nu)\rho}+K_{\mu(\nu\rho)})-\frac{1}{3}\,g_{\mu\nu}\,K^{\sigma}{_{\sigma\rho}}
+\frac{1}{3}\,g_{\nu\rho}\,K^{\sigma}{_{\sigma\mu}}~.
\end{align}
\end{subequations}
For completeness, the inverse relation is
\begin{equation}
K_{\mu\nu\rho}=
\breve{K}_{\mu\nu\rho}
+\hat{K}_{\mu\nu\rho}
+\frac{1}{3}\,g_{\mu\nu}\,\bar{K}_\rho
-\frac{1}{3}\,g_{\nu\rho}\,\bar{K}_\mu~.
\end{equation}
Upon applying the sigma matrices to the right hand side of \eqref{eq:ContorsionTensorComponents}, we obtain\footnote{The identity $\epsilon_{A[B}\epsilon_{CD]}=0$ is needed to simplify the result of the calculation.}
\begin{subequations}\label{eq:componentsofKtoSpinors}
\begin{align}
\begin{split}
&\breve{K}_{\mu\nu\rho}\
\to \
(\epsilon_{AC} \epsilon_{BF} \depsilon_{\dot{A} \dot{F}} \
\depsilon_{\dot{B} \dot{C}} -  \epsilon_{AF} \epsilon_{BC} \depsilon_{\dot{A} \dot{C}} \depsilon_{\dot{B} \dot{F}})(\Xi{}^{F\dot{F}} -  \tilde{\Xi}{}^{\dot{F} F})
\\
&\qquad\ \qquad =-i\, \varepsilon_{ABCF\Ad\Bd\Cd\Fd}\,(\Xi{}^{F\dot{F}} -  \tilde{\Xi}{}^{\dot{F} F})
\end{split}\label{eq:antisymmK}
~,\\[1em]
&\bar{K}_\mu \
\to\ 
3(\Xi_{A\Ad} + \tilde{\Xi}_{\Ad A})
~,\\[1em]
&\hat{K}_{\mu\nu\rho}\ 
\to \ 
\depsilon_{\dot{A} \dot{C}}\, \Omega_{ABC\dot{B}} + \
\epsilon_{AC}\, \tilde{\Omega}{}_{\dot{A} \dot{B} \dot{C} B}~.
\end{align}
\end{subequations}
The factor of $i$ in \eqref{eq:antisymmK} appears in Lorentzian signature only. The rest of this section is signature agnostic. 
Table \ref{tab:dofContorsion} summarises the share of degrees of freedom among the different tensor and spinor components. Under the map \eqref{eq:ContorsionasNSNS}, $\Omega_{ABC\Cd}=\tilde{\Omega}_{\Ad\Bd\Cd B}=0$, $\Xi_{A\Ad} + \tilde{\Xi}_{\Ad A}$ is related to $\partial_\mu \phi$ and $\Xi_{A\Ad} - \tilde{\Xi}_{\Ad A}$ to $\partial_\mu\sigma$.

\begin{table}
\centering
\begin{tabular}{llcr}
\toprule
&Tensor components & Spinor components & \hspace{2cm}d.o.f\\ \midrule
Antisymmetric&$\breve{K}_{\mu\nu\rho}$ & $\Xi_{A\Ad}-\tilde{\Xi}_{\Ad A}$& 4\\
Trace&$\bar{K}_{\mu}$ & $\Xi_{\,A\Ad}+\tilde{\Xi}_{\Ad A}$& 4\\
Traceless&$\hat{K}_{\mu\nu\rho}$ & $\Omega_{ABC\Ad}$~,\ \  $\tilde{\Omega}_{\Ad\Bd\Cd A}$& 16\\
\bottomrule
\end{tabular}
\caption{d.o.f in the different components of the contorsion.}\label{tab:dofContorsion}
\end{table}

\subsection{Riemann spinors}
A similar process can be followed to decompose the spinorial equivalent of $\mathfrak{R}_{\mu\nu\rho\sigma}$. 
First, we will exhaust the --- now smaller --- symmetry group of the tensor to identify its irreducible spinor components. 
Then, by means of \eqref{eq:FrakRasRandK}, we will relate the newly found spinors to the contorsion spinors and the usual curvature spinors of $R_{\mu\nu\rho\sigma}$.

We begin the spinoral reduction of the generalised Riemann tensor by implementing its only symmetries: the antisymmetry of both pairs of indices 
\begin{equation}
\begin{aligned}
\mathfrak{R}_{A\Ad B\Bd C\Cd D\Dd}
&=
\mathbf{X}_{ABCD}\depsilon_{\Ad\Bd}\depsilon_{\Cd\Dd}
+\tilde{\mathbf{X}}_{\Ad\Bd\Cd\Dd}\epsilon_{AB}\epsilon_{CD}
\\
&\hspace{1cm}
+\mathbf{\Phi}_{AB\Cd\Dd}\depsilon_{\smash{\Ad\Bd}}\epsilon_{CD}
+\tilde{\mathbf{\Phi}}_{\Ad\Bd CD}\epsilon_{AB}\depsilon_{\smash{\Cd\Dd}}~.
\end{aligned}
\end{equation}
The curvature spinors of $\mathfrak{R}$ are printed in bold typeface in order to distinguish them from those of $R$. 
Recall that $\mathfrak{R}_{\mu\nu\rho\sigma}\neq\mathfrak{R}_{\rho\sigma\mu\nu}$. 
The lack of this symmetry implies that $\mathbf{X}_{ABCD}\neq \mathbf{X}_{CDAB}$ and $\mathbf{\Phi}_{AB\Cd\Dd}\neq \tilde{\mathbf{\Phi}}_{\Cd\Dd AB}$ in general. 
The spinor $\mathbf{X}_{ABCD}$ is not completely symmetric and must be further reduced
\begin{equation}
\begin{aligned}
\mathbf{X}_{ABCD}=\mathbf{\Psi}_{ABCD}
-\big(
\mathbf{\Sigma}_{A(C}\epsilon_{D)B}
+\mathbf{\Sigma}_{B(C}\epsilon_{D)A}
\big)
+\mathbf{\Lambda}(\epsilon_{AC}\epsilon_{BD}+\epsilon_{AD}\epsilon_{BC})
~,
\end{aligned}
\end{equation}
where $\mathbf{\Psi}_{ABCD}$ and $\mathbf{\Sigma}_{AB}$ are completely symmetric. Putting tildes and dots yields the analogous expression for $\tilde{\mathbf{X}}_{\Ad\Bd\Cd\Dd}$. 
It might be worth remarking that $\tilde{\mathbf{\Lambda}}\neq\mathbf{\Lambda}$, because $\mathfrak{R}_{\mu[\nu\rho\sigma]}\neq 0$. 
All the remaining spinors are completely symmetric and hence irreducible. 
Table \ref{tab:curvaturedof}  shows how the degrees of freedom encoded in the irreducible spinors add up to $36$, the number of independent (real) components of $\mathfrak{R}_{\mu\nu\rho\sigma}$ \cite{Penrose:1983mf}. 
These degrees of freedom also include the Ricci tensor and the Ricci scalar, which can be obtained from the same spinor components
\begin{align}
	\mathfrak{R}_{\mu\rho\nu}{}^\rho 
	&\to
	-\mathbf{\Phi}_{AB\Ad\Bd}
	-\tilde{\mathbf{\Phi}}_{\Ad\Bd AB}
	+4(\mathbf{\Sigma}_{AB}\depsilon_{\smash{\Ad\Bd}}
		+\tilde{\mathbf{\Sigma}}_{\smash{\Ad\Bd}}	\epsilon_{AD})
	+3(\mathbf{\Lambda}+\tilde{\mathbf{\Lambda}})\epsilon_{AB}\depsilon_{\smash{\Ad\Bd}}
	\\
	\mathfrak{R}_{\mu\nu}{}^{\mu\nu} 
	&=
	12(\mathbf{\Lambda}+\tilde{\mathbf{\Lambda}})~.
\end{align}

\begin{table}
\centering
\begin{tabular}{lcc}
\toprule
Spinor component &  $\mathfrak{R}_{\mu\nu\rho\sigma}$ & $R_{\mu\nu\rho\sigma}$\\ \midrule
$\Psi$, $\tilde \Psi$&$\qquad 2\times 5\qquad$ &$\qquad 2\times 5\qquad $ \\
$\Sigma$, $\tilde\Sigma$&$2\times 4$& $0$ \\
$\Lambda$, $\tilde\Lambda$&$2\times 1$& $1$\\
$\Phi$, $\tilde\Phi$ &$2\times 9$&$9$\\ \midrule
Total &$38$ & $20$\\
\bottomrule
\end{tabular}
\caption{d.o.f counting for the curvature spinors.}\label{tab:curvaturedof}
\end{table}

The inverse spinorial identities
\begin{align}
&\mathbf{\Phi}_{AB\Cd\Dd}=\frac{1}{4}\mathfrak{R}_{A\Ad B}{}^{\Ad}{}_{C\Cd}{}^{C}{}_{\Dd}~,\label{eq:PhiasfrakR}\\[0.7em]
&\mathbf{X}_{ABCD}=\frac{1}{4}\mathfrak{R}_{A\Ad B}{}^{\Ad}{}_{C\Cd D}{}^{\Cd}~,\\[0.7em]
&\mathbf{\Psi}_{ABCD}=\frac{1}{4!}\mathbf{X}_{(ABCD)}~,\\[0.7em]
&\mathbf{\Sigma}_{AB}=\frac{1}{8}\mathbf{X}_{(A|C|B)}{}^C~,\\[0.7em]
&\mathbf{\Lambda}=\frac{1}{6}\mathbf{X}_{AB}{}^{AB}~,\label{eq:LambdaasX}
\end{align}
are better suited expressions for computing the spinors of a given solution.

So far, we have identified the irreducible parts that make up the contorsion and the Riemann tensor. 
However, the Riemann tensor and the contorsion are not independent. 
Their relation is made explicit in equation \eqref{eq:FrakRasRandK}. 
Hence, the bold curvature spinors must be functions of the contorsion spinors and the usual curvature spinors. 
The procedure to establish these relations is straightforward. 
First, the spinorial equivalent of the right hand side of \eqref{eq:FrakRasRandK} must be obtained. 
Then, applying \eqref{eq:PhiasfrakR}--\eqref{eq:LambdaasX} yields the desired expressions:
\begin{align}
	\mathbf{\Phi}_{AB\Cd\Dd}
	&=
	\Phi_{AB\Cd\Dd}
	+\frac{1}{8}\,\nabla_{(A|\Ad }\tilde{\Theta}_{\Cd}{}^{\Ad}{}_{\Dd |B)}
	-\frac{1}{8}\tilde{\Theta}_{\Cd}{}^{\Ad\Ed}{}_{(A|}\,\tilde{\Theta}_{\Dd\Ad\Ed|B) }
	\nonumber\\[1em]
	\begin{split}
		&= 
		\Phi_{AB\Cd\Dd}
		+\frac{1}{2}\bigg(\nabla_{(A|\Ad }\tilde{\Omega}_{\Cd}{}^{\Ad}{}_{\Dd |B)}
		+\nabla_{A(\Cd }\tilde{\Xi}_{ \Dd)B}
		+\nabla_{B(\Cd }\tilde{\Xi}_{\Dd)A}
		\\
		&\hspace{2cm}
		-4\,\tilde{\Xi}_{\Cd (A|}\tilde{\Xi}_{\Dd |B)}
		-2\,\tilde{\Xi}{}^{\Ad}{}_{(A|}\tilde{\Omega}_{\Cd\Dd\Ad |B)}
		-\tilde{\Omega}_{\Cd}{}^{\Ad\Ed}{}_{(A|}\tilde{\Omega}_{\Dd \Ad \Ed |B)}\bigg)~,
	\end{split}\label{eq:frakPhiaskappa12}
\end{align}
\begin{align}
	\mathbf{\Psi}_{ABCD}
	&=
	\Psi_{ABCD}
	+\frac{1}{4!}\bigg(\nabla_{\Ad(A}\Theta_{BCD)}{}^{\Ad}
	-\Theta_{(AB}{}^{E\Ad}\Theta_{CD)E\Ad}\bigg)
	\nonumber\\[0.5em]
	\begin{split}
		&= 
		\Psi_{ABCD}
		+\frac{1}{4!}\bigg(\nabla_{\Ad (A }\Omega_{BCD)}{}^{\Ad}
		+2\,\Xi_{(A}{}^{\Ad}\,\Omega_{BCD)\Ad}
		-\Omega_{(AB}{}^{E\Ad}\,\Omega_{CD)E\Ad}\bigg)\,\,
	\end{split}
\end{align}
\begin{align}
	\mathbf{\Sigma}_{AB}
	&=
	-\frac{1}{4}\nabla_{\Ad (A}\Xi_{B)}{}^{\Ad}
	+\frac{1}{8}\nabla_{C\Ad}\,\Omega_{AB}{}^{C\Ad}
	+\frac{3}{4}\,\Xi^{\ C\Ad}\,\Omega_{ABC\Ad}~,
\\[2.5em]
	\mathbf{\Lambda}
	&=
	\Lambda
	+\frac{1}{6}\,	\nabla_{B\Ad}\Theta^A{}_{A}{}^{B\Ad}
	+\frac{1}{12}\, \Theta_{ABC\Ad}\ \Theta^{ABC\Ad}
	+\frac{1}{12}\, \Theta^A{}_{A}{}^{B\Ad}\ \Theta_B{}^E{}_{E\Ad}
	\nonumber\\[0.5em]
	&=
	\Lambda
	-\frac{1}{2}\nabla^{A\Ad}\Xi_{A\Ad}
	-\Xi_{A\Ad}\ \Xi{}^{A\Ad}
	+\frac{1}{12}\, \Omega_{ABC\Ad}\ \Omega{}^{ABC\Ad}~.\label{eq:frakLambdakappa12}
\end{align}

\bibliography{bibl}
\bibliographystyle{JHEP}

\end{document}